\documentclass[a4paper,USenglish,cleveref,autoref,numberwithinsect,thm-restate]{lipics-v2021}

\usepackage{amsmath, amssymb, amstext, amsthm, mathtools, bm}
\usepackage{cancel}
\usepackage{xspace}
\usepackage{stmaryrd}
\usepackage[numbers,square,comma,longnamesfirst,sort]{natbib}
\usepackage[procnumbered, linesnumbered,ruled]{algorithm2e}

\SetCommentSty{mycommfont}
\usepackage{framed}
\usepackage[framemethod=TikZ]{mdframed}

\usepackage{todonotes}

\numberwithin{equation}{section}
\numberwithin{figure}{section}

\Crefname{claim}{Claim}{Claims}

\newcommand{\Nat}{\mathbb{N}} % Set of natural numbers including 0
\newcommand{\nat}{\mathbb{N}} % Set of natural numbers including 0
\newcommand{\Int}{\mathbb{Z}} % Set of integer numbers
\newcommand{\deff}{\coloneqq}
\newcommand{\from}{\colon}% Colon for definition of functions

\newcommand{\floor}[1]{{\left\lfloor #1 \right\rfloor}}

\newcommand{\cost}{\mathrm{cost}}
\DeclarePairedDelimiter{\abs}{\lvert}{\rvert}
\DeclarePairedDelimiter{\iverson}{\llbracket}{\rrbracket}
\newcommand{\conv}{\mathbin{\odot}}
\newcommand{\fconv}[1][f]{\mathbin{\circledast_{#1}}} % Convolution operator with function "f" as default function
\newcommand{\Mat}{\mathsf{Mat}}

\newcommand{\fConv}[1][f]{\textnormal{\ensuremath{#1}-\textsc{Convolution}}\xspace}

\newcommand{\fquery}[1][f]{\textnormal{\ensuremath{#1}-\textsc{Query}}\xspace}

\newcommand{\CyConv}{\textnormal{\textsc{Cyclic Convolution}}\xspace}
\newcommand{\CyConvProb}{\textnormal{\textsc{Cyclic Convolution Problem}}\xspace}

\newcommand{\Mrange}{{\{-M,\ldots, M\}}}

\DeclareMathOperator{\polylog}{\mathrm{polylog}}

\newcommand{\ZZ}{{\mathbb Z}}
\newcommand{\FF}{{\mathbb F}}

\newcommand{\Oh}{\mathcal{O}}
\newcommand{\Ot}{\widetilde{\mathcal{O}}}

\newcommand{\Oc}[1]{\Ot(#1 \cdot \mathrm{polylog}(M))}
\newcommand{\Occ}[1]{\Ot\left(#1 \cdot \mathrm{polylog}(M)\right)}

\newcommand{\indeg}{\mathrm{indeg}} %In-degree of a vertex
\newcommand{\outdeg}{\mathrm{outdeg}} %Out-degree of a vertex

\newcommand{\Vsrc}{V_{\textup{\textsf{src}}}} % Source vertices
\newcommand{\Vmid}{V_{\textup{\textsf{mid}}}} % Middle vertices
\newcommand{\Vsnk}{V_{\textup{\textsf{snk}}}} % Sink vertices

\newcommand{\defi}{\mathrm{defi}} % deficiency of a vertex
\newcommand{\Dall}{\mathrm{Defi}}%_{\!\textsf{all}}} % Total deficiency of a graph

\newcommand{\Pot}{\mathrm{LHS}} % Potential

\newcommand{\Dd}{D}%{\mathcal{D}}
\newcommand{\Part}{\mathcal{P}} %Notation for the cyclic partition

\let\eps\epsilon

\newcommand{\sigmaL}{\bm{\sigma}_{\type}^{L}}
\newcommand{\sigmaR}{\bm{\sigma}_{\type}^{R}}
\newcommand{\sigmaT}{\bm{\sigma}_{\type}^{T}}

\newcommand{\domL}{L_{\type}}
\newcommand{\domR}{R_{\type}}

\newcommand{\domZ}{Z_{\type}}

\newcommand{\Lone}{\textnormal{L}^{\high}}
\newcommand{\Ltwo}{\textnormal{L}^{\low}}
\newcommand{\Rone}{\textnormal{R}^{\high}}
\newcommand{\Rtwo}{\textnormal{R}^{\low}}

\newcommand{\tr}{\textnormal{tr}}
\newcommand{\ba}{\mathbf{a}}
\newcommand{\bb}{\mathbf{b}}

\newcommand{\bv}{\mathbf{v}}
\newcommand{\bu}{\mathbf{u}}
\newcommand{\bx}{\mathbf{x}}
\newcommand{\by}{\mathbf{y}}
\newcommand{\bz}{\mathbf{z}}
\newcommand{\bw}{\mathbf{w}}
\newcommand{\bq}{\mathbf{q}}
\newcommand{\br}{\mathbf{r}}
\newcommand{\bs}{\mathbf{s}}
\newcommand{\bt}{\mathbf{t}}
\newcommand{\high}{{(\mathrm{high})}}
\newcommand{\low}{{(\mathrm{low})}} 
\newcommand{\type}{{\overline{p}}}
\newcommand{\DP}{{\mathsf{DP}}}

\renewcommand{\prime}{{\textsf{prime}}}
\newcommand{\primebound}{{\textsf{prime\_bound}}}

\newcounter{openquestion}
\newenvironment{openquestion}
{\mdfsetup{%
    nobreak=true,
	middlelinecolor=gray,
	middlelinewidth=1pt,
	backgroundcolor=gray!10,
    innertopmargin=7pt,
	roundcorner=5pt}
\begin{mdframed}\refstepcounter{openquestion}
\textbf{Question \theopenquestion:}}
{\end{mdframed}}
\newenvironment{insight}
{\mdfsetup{%
    nobreak=true,
	middlelinecolor=gray,
	middlelinewidth=1pt,
	backgroundcolor=gray!10,
    innertopmargin=7pt,
	roundcorner=5pt}
\begin{mdframed}}
{\end{mdframed}}

\title{Computing Generalized Convolutions\\ Faster Than Brute Force}
\titlerunning{Computing Generalized Convolutions Faster Than Brute Force}
\author{Bar\i\c{s} Can Esmer}
  {CISPA Helmholtz Center for Information Security, Germany}
  {baris-can.esmer@cispa.de}
  {https://orcid.org/0000-0001-5694-1465}{}
\author{Ariel Kulik}
  {CISPA Helmholtz Center for Information Security, Germany}
  {ariel.kulik@cispa.de}
  {}{}
\author{D\'aniel Marx}
  {CISPA Helmholtz Center for Information Security, Germany}
  {marx@cispa.de}
  {https://orcid.org/0000-0002-5686-8314}{}
\author{Philipp Schepper}
  {CISPA Helmholtz Center for Information Security, Germany}
  {philipp.schepper@cispa.de}
  {https://orcid.org/0000-0002-5810-7949}{}
\author{Karol W\k{e}grzycki}
  {Saarland University and Max Planck Institute for Informatics, Saarbr\"ucken, Germany}
  {wegrzycki@cs.uni-saarland.de}
  {https://orcid.org/0000-0001-9746-5733}
  {}

\authorrunning{B.\,C.~Esmer, A.~Kulik, D.~Marx, P.~Schepper, and K.~W\k{e}grzycki}
\Copyright{Bar\i\c{s} Can Esmer, Ariel Kulik, D\'aniel Marx, Philipp Schepper, and Karol W\k{e}grzycki}

\funding{%
Research supported by the European Research Council (ERC) consolidator grant No.~725978 SYSTEMATICGRAPH and the project TIPEA (grant No. 850979).
}

\acknowledgements{%
We would like to thank Karl Bringmann and Jesper Nederlof for useful discussions.
Bar\i\c{s} Can Esmer and Philipp Schepper are part of Saarbrücken Graduate School of Computer Science, Germany.
}



\nolinenumbers
\hideLIPIcs

\begin{document}

\maketitle

\begin{abstract}
In this paper, we consider a general notion of convolution.
Let $\Dd$ be a finite domain and let $\Dd^n$ be the set of $n$-length vectors
(tuples) of $\Dd$. Let $f \from \Dd \times \Dd \to \Dd$ be a function and
let $\oplus_f$ be a coordinate-wise application of $f$. The \fConv of two
functions $g,h \from \Dd^n \to \{-M,\ldots,M\}$ is 
\begin{displaymath}
(g \fconv h)(\bv) \deff \sum_{\substack{\bv_g,\bv_h \in
		\Dd^n\\ \text{s.t. } \bv = \bv_g \oplus_f \bv_h}} g(\bv_g) \cdot h(\bv_h)
\end{displaymath}
for every $\bv \in \Dd^n$.
This problem generalizes many fundamental convolutions
such as Subset Convolution, XOR Product, Covering Product or Packing Product,
etc. 
For arbitrary function $f$ and domain $\Dd$ we can compute \fConv via brute-force enumeration
in $\Oc{|\Dd|^{2n}}$ time.

Our main result is an improvement over this naive algorithm. We show that \fConv
can be computed exactly in $\Oc{ (c \cdot |\Dd|^2)^{n}}$ for constant $c
\deff 3/4$ when $\Dd$ has even cardinality.
Our main observation is that a
\emph{cyclic partition} of a function $f \from \Dd \times \Dd \to \Dd$ can
be used to speed up the computation of \fConv, and we show that an appropriate
cyclic partition exists for every $f$.

Furthermore, we demonstrate that a single entry of the \fConv can be computed
more efficiently. In this variant, we are given two functions $g,h \from \Dd^n
\to \{-M,\ldots,M\}$ alongside with a vector $\bv \in \Dd^n$ and the task of
the \fquery problem is to compute integer $(g \fconv h)(\bv)$. This is a
generalization of the well-known Orthogonal Vectors problem. We show that
\fquery can be computed in $\Oc{|\Dd|^{\frac{\omega}{2} n}}$ time, where $\omega
\in [2,2.372)$ is the exponent of currently fastest matrix multiplication
algorithm.

\end{abstract}

% \todo[inline]{PS: We should change $\sigma_C$ to something like $\overline \sigma_C$ or $\tau_C$ to make clear it is not going in the same direction as $\sigma_A$ and $\sigma_B$.\\
% AK: I suggest we implement this change in a journal version. I do not think it worths the effort at the moment.}
% \ippp{Should we make $\type$ bold and omit the bar?}

% \clearpage
% \setcounter{page}{1}

\section{Introduction}

Convolutions occur naturally in many algorithmic applications, especially in the
exact and parameterized algorithms. The most prominent example is a subset
convolution procedure~\cite{hall1934contribution,weisner1935abstract}, for which
an  efficient $\Oc{2^n}$ time algorithm for subset convolution dates back to
Yates~\cite{yates1937design} but in the context of exact algorithms it was first
used by Bj\"orklund et al.~\cite{fast-subset-convolution}.\footnote{We use $\Ot(\cdot)$ notation to
hide polylogarithmic factors. We assume that $M$ is the maximum absolute value of the
integers on the input.}
Researchers considered a plethora of other variants of convolutions, such as:
Cover Product, XOR Product, Packing Product, Generalized Subset Convolution, or
Discriminantal Subset
Convolution~\cite{fast-subset-convolution,covering-packing-11,BjorklundHKK09,bjorklund-06,Rooij09,generalized-conv-cygan,brand2022}.
These subroutines are crucial ingredients in the design of efficient algorithms
for many exact and parameterized algorithms such as Hamiltonian Cycle, Feedback
Vertex Set, Steiner Tree, Connected Vertex Cover, Chromatic Number, Max $k$-Cut
or Bin
Packing~\cite{cut-and-count,bjorklund-06,zamir-icalp21,binpacking,bjorklund-parity,clifford-algebras}.
These convolutions are especially useful for dynamic programming algorithms on
tree decompositions and occur naturally during join operations (e.g.,
\cite{Rooij09,cut-and-count,csr21}).  Usually, in the process of algorithm
design, the researcher needs to design a different type of convolution from
scratch to solve each of these problems. Often this is a highly technical and
laborious task. Ideally, we would like to have a single tool that can be used as
a blackbox in all of these cases. This motivates the following ambitious goal in
this paper:

\begin{insight}
    \textbf{Goal:} Unify convolution procedures under one general umbrella.
\end{insight}
Towards this goal, we consider the problem of computing $f$-\emph{Generalized
Convolution} (\fConv) introduced by van Rooij~\cite{csr21}. Let $\Dd$ be a
finite domain and let $\Dd^n$ be the $n$ length vectors (tuples) of $\Dd$. Let
$f \from \Dd \times \Dd \to \Dd$ be an arbitrary function and let $\oplus_f$ be
a coordinate-wise application of the function $f$.\footnote{We provide a formal
definition of $\oplus_f$ in Section~\ref{sec:prelim}.}
For two functions $g,h \from \Dd^n \to \Int$
the \fConv, denoted by $(g \fconv h) \from \Dd^n \to \Int$,
is defined for all $\bv \in \Dd^n$ as
\begin{displaymath}
    (g \fconv h)(\bv) \deff \sum_{\substack{\bv_g,\bv_h \in
    \Dd^n\\ \text{s.t. } \bv = \bv_g \oplus_f \bv_h}} g(\bv_g) \cdot h(\bv_h)
    .
\end{displaymath}
Here we consider a standard $\Int(+,\cdot)$
ring. Through the paper we assume that $M$ is the absolute value of the maximum
integer given on the input.

In the \fConv problem the functions $g,h \from D^n \to \Mrange$ are given as an input
and the output is the function $(g \fconv h)$. Note, that
the input and output of the \fConv problem consist of
$3\cdot |\Dd|^n$ integers. Hence it is conceivable that \fConv could be solved
in $\Oc{|\Dd|^n}$. Such a result for arbitrary $f$ would be a real
breakthrough in how we design parameterized algorithms. 
So far, however, researchers have focused on
characterizing functions $f$ for which
\fConv can be solved in $\Oc{|\Dd|^n}$ time. 
In~\cite{csr21}  van Rooij considered specific instances of this  setting, where for some constant
$r \in \nat$ the function $f$ is defined as either (i) standard
addition: $f(x,y) \deff x+y$, or (ii) addition with a maximum: $f(x,
y) \deff \min(x+y,r-1)$, or (iii) addition modulo $r$, or
(iv) maximum: $f(x,y) \deff \max(x,y)$. Van Rooij~\cite{csr21} showed that for
these special cases the \fConv can be solved in $\Oc{|\Dd|^n}$
time. His results allow the function $f$ to differ between coordinates.
A recent result regarding generalized Discrete Fourier Transform~\cite{umans19}
can be used  in conjunction with Yates algorithm~\cite{yates1937design} to
compute \fConv in $\Oc{|\Dd|^{\omega \cdot n / 2}}$ time when
$f$ is a finite-group operation and $\omega$ is the exponent of the
currently fastest matrix-multiplication algorithms.\footnote{This observation was brought to our attention by Jesper Nederlof~\cite{jesper}.} 
To the best of our knowledge
these are the most general settings where convolution has been considered so
far. 

Nevertheless, for an arbitrary function $f$, to the best of our knowledge the
state-of-the-art for \fConv is a straightforward quadratic time
enumeration.

\begin{openquestion}
    \label{open:main}
    Is the naive $\Oc{|\Dd|^{2n}}$ algorithm for \fConv optimal?
\end{openquestion}
Similar questions were studied from the point of view of the Fine-Grained
Complexity. In that setting the focus is on convolutions with sparse
representations, where the input size is only the size of the support of the
functions $g$ and $h$. It is conjectured that even subquadratic algorithms are
highly unlikely for these representations~\cite{cygan-talg19,paturi-icalp17}.
However, these lower bounds do not answer Question~\ref{open:main}, because
they are highly dependent on the sparsity of the input.

\subsection{Our Results}

We provide a positive answer to Question~\ref{open:main}
and show an exponential improvement (in $n$) over a naive
$\Oc{|\Dd|^{2n}}$ algorithm for every function $f$. 

\begin{theorem}[Generalized Convolution]
    \label{thm:main-thm}
    Let $\Dd$ be a finite set and $f\from \Dd\times \Dd \to \Dd$.
    There is an algorithm for \fConv with the following running time
            $\Occ{ \left(
                    \frac{3}{4} \cdot |\Dd|^2
            \right)^{n} }
            $ when $\abs{D}$ is even, or
            $\Occ{ \left(
                    \frac{3}{4} \cdot |\Dd|^2+\frac{1}{4}\cdot|\Dd|
            \right)^{n} }$ when $\abs{D}$ is odd.
\end{theorem}
Observe that the running time obtained by Theorem~\ref{thm:main-thm} improves
upon the brute-force for every $|\Dd| \ge 2$.
Our technique
works in a more general setting when $g \from L^n \to \Int$
and $h\from R^n \to \Int$ and $f\from L \times R \to T$ for arbitrary
domains $L,R$ and $T$ (see Section~\ref{sec:prelim} for the exact
running time dependence).

\subparagraph*{Our Technique: Cyclic Partition.}

Now, we briefly sketch the idea behind the proof of Theorem~\ref{thm:main-thm}.
We say that a function is $k$-cyclic if it can be represented as an addition
modulo $k$ (after relabeling the entries of the domain and image). These
functions are somehow simple, because as observed in~\cite{csr21,Rooij20} \fConv
can be computed in $\Oc{k^n}$ time if $f$ is $k$-cyclic. In a nutshell, our idea
is to partition the function $f\from \Dd \times \Dd \to \Dd$ into
cyclic functions and compute the convolution on these parts independently.

\begin{figure}[ht!]
    \centering
    \includegraphics[width=0.7\textwidth]{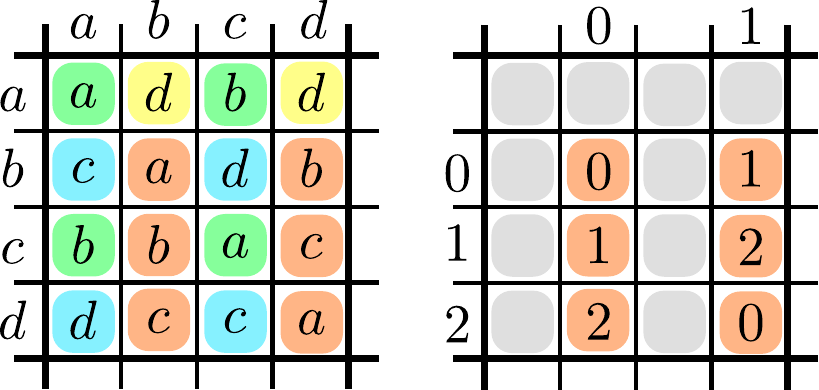}
    \caption{Left figure illustrates exemplar function $f \from \Dd \times \Dd \to
    \Dd$ over domain $\Dd \deff \{a,b,c,d\}$. We highlighted a cyclic
    partition with red, blue, yellow and blue colors. Each color represents a
    different minor of $f$. On the right figure we
    demonstrate that the red-highlighted minor can be represented as addition modulo $3$ (after
    relabeling $a \mapsto 0$, $b\mapsto 1$ and $c \mapsto 2$). Hence
    the red minor has cost $3$. The reader can further verify that green and blue minors have
    cost $2$ and yellow minor has cost $1$, hence the cost of that particular
    partition is $3+2+2+1 = 8$.}
    \label{fig:cyclic-partition}
\end{figure}

More formally, a cyclic minor
of the function $f\from \Dd \times \Dd \to \Dd$ is a (combinatorial) rectangle
$A \times B$ with $A,B\subseteq
\Dd$ and a number $k\in \Nat$ such that $f$ restricted to $A,B$ is a $k$-cyclic function. The cost of the cyclic minor
 $(A,B,k)$ is
$ \cost (A,B) \deff k $. A cyclic
partition is a set $\{(A_1,B_1,k_1),\ldots,(A_m,B_m,k_m)\}$ of cyclic minors
such that for every $(a,b)
\in \Dd \times \Dd$ there exists a unique $i \in [m]$ with $(a,b) \in A_i \times
B_i$. The cost of the cyclic partition $\Part =
\{(A_1,B_1,k_1),\ldots,(A_m,B_m,k_m)\}$ is $\cost(\Part) \deff \sum_{i=1}^m
k_i$. See Figure~\ref{fig:cyclic-partition} for an example of a cyclic partition.

Our first technical contribution is an algorithm to compute \fConv when the cost of a cyclic partition is small.

\begin{lemma}[Algorithm for \fConv]
    \label{lem:alg}
    Let $\Dd$ be an arbitrary finite set, $f\from \Dd \times \Dd \to
    \Dd$ and let $\Part$ be the cyclic partition of $f$. Then there exists
    an algorithm which given $g,h\from \Dd^n \to \Int$ computes
    $(g \fconv h)$ in $\Oc{(\cost(\Part)^n + |\Dd|^n)}$ time.
\end{lemma}
The idea behind the proof of Lemma~\ref{lem:alg} is as follows. Based on the partition
$\Part$,  for any pair of vectors $\bu,\bw \in \Dd^n$,
we can define a type $\type \in
[m]^n$ such that $(\bu_i,\bw_i) \in A_{\type_i} \times B_{\type_i}$
for every $i \in [n]$. Our main idea is to go over each type $\type$ and
compute the sum in the definition of \fConv only for pairs $(\bv_g,\bv_h)$ that
have type $\type$. In order to do this, first we select the vectors $\bv_g$
and $\bv_h$ that are compatible with this type $\type$. For instance, consider
the example in Figure~\ref{fig:cyclic-partition}. Whenever $\type_i$ refers to,
say, the red-colored minor, then we consider $\bv_g$ only if its $i$-th coordinate is in $\{b,c,d\}$
and consider $\bv_h$ only if its $i$-th coordinate is in $\{b,d\}$. After computing all
these vectors $\bv_g$ and $\bv_h$, we can transform them according to the cyclic
minor at each coordinate. Continuing our example, as the red-colored minor is
$3$-cyclic, we can represent the $i$-th coordinate of $\bv_g$ and $\bv_h$ as
$\{0,1,2\}$ and then the problem reduces to addition modulo $3$ at that coordinate.  Therefore, using
the algorithm of van Rooij~\cite{csr21} for cyclic convolution we can handle all
pairs of type $\type$ in $\Oc{(\prod_{i=1}^n k_{\type_i})}$ time. 
As we go over all $m^n$ types $\type$ the sum of $m^n$ terms is
\begin{displaymath}
  \sum_{\type \in [m]^n} \left(\prod_{i=1}^n
  k_{\type_i}\right) = \left(\sum_{i=1}^m k_i\right)^n =
  \cost(\Part)^n
   .
\end{displaymath}

Hence, the overall running time is $\Oc{\cost(\Part)^n}$. This running time evaluation
ignores the generation of the vectors given as input for the cyclic convolution
algorithm. The efficient computation of these vectors is nontrivial and requires
further techniques that we explain in Section~\ref{sec:algo}.

It remains to provide the low-cost cyclic partition of an arbitrary function $f$.

\begin{lemma}
  \label{lem:cost-bound}
  For any finite set $\Dd$ and any function $f\from \Dd \times \Dd
  \to \Dd$ there is a cyclic partition $\Part$ of $f$
  such that $\cost(\Part) \le \frac{3}{4} |\Dd|^2$ when $\abs{\Dd}$ is even, or 
  $\cost(\Part) \le \frac{3}{4} |\Dd|^2 + \frac{1}{4} |\Dd|$ when $\abs{\Dd}$ is
  odd.
\end{lemma}
For the sake of presentation let us assume that $|\Dd|$ is even.
In order to show Lemma~\ref{lem:cost-bound}, we partition $\Dd$ into pairs
$A_1,\ldots,A_{k}$ where $k \deff |\Dd|/2$ and consider the
restrictions of $f$ to $A_j \times \Dd$ one by one.
Intuitively, we partition the $\Dd \times \Dd$ table describing $f$ into pairs
of rows and give a bound on the cost of each pair.
This partition allows us to encode $f$ on $A_j \times \Dd$ as a directed graph $G$ with $|\Dd|$
edges and $|\Dd|$ vertices.
We observe that directed cycles and directed paths can be represented as
cyclic minors. Our goal is to partition graph $G$ into such subgraphs in a way that the total cost of the resulting cyclic
partition is small.
Following this argument, 
the proof of Lemma~\ref{lem:cost-bound}  becomes a graph-theoretic analysis. 
The proof of Lemma~\ref{lem:cost-bound} is
included in Section~\ref{sec:existenceMinor}. 
We also give an example which suggests that the constant $\frac{3}{4}$ in~\cref{lem:cost-bound} cannot be improved further
while using the partition of $\Dd$ into arbitrary pairs
(see~\cref{lem:existenceOfMinor:tightness}).

Our method applies for more
general functions $f \from L \times R \to T$, where
domains $L,R,T$ can be different and have arbitrary cardinality.
We note that a weaker variant of Lemma~\ref{lem:cost-bound}
in which the guarantee is $\cost(\Part_f) \leq \frac{7}{8} |\Dd|^2$ is easier to
attain (see Section~\ref{sec:existenceMinor}).

% Finally, we complement the upperbound in~\cref{lem:cost-bound} with an example of
% a function $f : L \times R \to T$ with $|L| = 2$ for which the cyclic partition
% returned by our method is optimal (see~\cref{lem:existenceOfMinor:tightness}).
% \ariel{try to mention the $\frac{3}{4}$ factor is optimal if we consider two rows at a time}

\subparagraph*{Efficient Algorithm for Convolution Query.}

Our next contribution is an efficient algorithm to query a single value of
\fConv. In the $\fquery$ problem,  the input is  $g,h \from \Dd^n \to
\Int$ and a single vector $\bv \in \Dd^n$. The task is to compute a value
$(g \fconv h)(\bv)$. Observe that this task generalizes\footnote{It is a special
case with $\Dd = \{0,1\}$, $\bv = 0^n$ and $f(x,y) = x \cdot y$} the fundamental
problem of Orthogonal Vectors.  We show that computing $\fquery$ is much faster
than computing the full output of $\fConv$.

\begin{theorem}[Convolution Query]
  \label{thm:membership}
  For any finite set $\Dd$ and function $f\from \Dd \times \Dd \to \Dd$ there is a
  $\Oc{|\Dd|^{\omega \cdot n / 2} }$ time algorithm for the $\fquery$ problem.
\end{theorem}
Here $\Oc{m^\omega}$ is the time needed to multiply two $m \times m$ integer
matrices  with values in $\Mrange$ and currently $\omega \in
[2,2.372)$~\cite{mm-exponent,duan-fmm}. 
Note, that under the assumption that two matrices can be multiplied in the linear in the input
time (i.e., $\omega = 2$) then Theorem~\ref{thm:membership} runs in the nearly-optimal
$\Oc{|\Dd|^n}$ time. Theorem~\ref{thm:membership} is significantly faster
than Theorem~\ref{thm:main-thm} even when we plug-in the naive algorithm for matrix
multiplication (i.e., $\omega = 3$). The proof of Theorem~\ref{thm:membership}
is inspired by an interpretation of the $\fquery$  problem as counting length-$4$ cycles in a graph.

\subsection{Related Work}

Arguably, the problem of computing the Discrete Fourier Transform (DFT) is the
prime example of convolution-type problems in computer science. Cooley and
Tukey~\cite{fft} proposed the fast algorithm to compute DFT. Later, Beth 
\cite{beth1984verfahren} and Clausen~\cite{clausen1989fast} initiated the study
of generalized DFTs whose goal has been to obtain a fast algorithm for
DFT where the underlying group  is arbitrary.
After a long line of
works~(see~\cite{rockmore2004recent} for the survey), the currently best algorithm for
generalized DFT concerning group $G$ runs in $\Oh(|G|^{\omega/2+\eps})$
operations for every $\eps > 0$~\cite{umans19}.

A similar technique to ours was introduced by Bj\"orklund et
al.~\cite{zeta-on-lattices}. The paper gave a characterization of lattices that
admit a fast zeta transform and a fast M\"obius transform.  Their paper used the
notion of \emph{covering pairs}, which is similar to cyclic partitions used in
this paper but with a completely different goal.

From the lower-bounds perspective to the best of our knowledge only
a naive $\Omega(|\Dd|^n)$ lower bound is known for \fConv (as
this is the output size). We note that known lower bounds for different
convolution-type problems, such as
$(\min,+)$-convolution~\cite{cygan-talg19,paturi-icalp17},
$(\min,\max)$-convolution~\cite{bringmann2019approximating}, min-witness
convolution~\cite{min-witness}, convolution-3SUM~\cite{conv-3sum} or even skew-convolution~\cite{skew-conv}
cannot be easily adapted to $\fConv$ as the hardness of these problems comes
primarily from the ring operations.

The Orthogonal Vector problem is related to the \fquery problem. In the
Orthogonal Vector problem we are given two sets of $n$ vectors $A,B \subseteq
\{0,1\}^d$ and the task is to decide if there is a pair $a \in A$, $b \in B$
such that $a \cdot b = 0$.  In~\cite{ov-seth} it was shown that no $n^{2-\eps}
\cdot 2^{o(d)}$ algorithm for Orthogonal Vectors is possible for any $\eps > 0$
assuming SETH~\cite{icm-survey}. The currently best algorithm for Orthogonal
Vectors run in $n^{2-1/\Oh(\log(d)/\log(n))}$
time~\cite{polynomial-method2,polynomial-method1}, $\Oh(n \cdot 2^{cd})$ for
some constant $c < 0.5$~\cite{NederlofW21}, or $\Oh(|{\downarrow}A| +
|{\downarrow}B|)$~\cite{BjorklundHKK09} (where $|{\downarrow}F|$ is the total
number of vectors whose support is a subset of the support of input vectors).

\subsection{Organization}

In Section~\ref{sec:prelim} we provide the formal
definitions of the problems alongside the general statements of our results.
In Section~\ref{sec:algo} we give an algorithm for \fConv that uses
a given cyclic partition. In Section~\ref{sec:existenceMinor} we show that for
every function $f \from \Dd \times \Dd \to \Dd$ there exists a cyclic
partition of low cost. Finally, in Section~\ref{sec:gen-ov} we give an algorithm
for \fquery and prove Theorem~\ref{thm:membership}. In
Section~\ref{sec:conclusion} we conclude the paper and discuss future work.

\section{Preliminaries} \label{sec:prelim}

Throughout the paper, we use Iverson bracket notation, where for the logic expression
$P$, the value of $\iverson{P}$ is $1$ when $P$ is true and $0$ otherwise. For $n \in
\nat$ we use $[n]$ to denote $\{1,\ldots,n\}$. Through the paper we
denote vectors in bold, for example, $\bq \in \Int^k$ denotes a $k$-dimensional
vector of integers. We use subscripts to denote the entries of the vectors,
e.g., $\bq \deff (\bq_1,\ldots,\bq_k)$.

Let $L$, $R$ and $T$ be arbitrary sets
and let $f \from L \times R \to T$ be an arbitrary function.
We extend the definition of such an arbitrary function $f$ to
vectors as follows.  For two vectors $\bu \in L^n$ and $\bw\in R^n $ we define
\[
  \bu \oplus_f \bw \deff (f(\bu_1,\bw_1),\ldots, f(\bu_n,\bw_n)).
\]

In this paper, we consider the \fConv problem with a more general domain and image. We define it formally as follows:

\begin{definition}[$f$-Convolution]\label{def:fconv}

    Let $L$, $R$ and $T$ be arbitrary sets and let $f\from L \times R \to T$ be an arbitrary function.
    The \fConv of two functions $g\from L^n \to
  \Int$ and $h\from R^n\to \Int$, where $n\in \Nat$, is the
  function $(g \fconv h)\from T^n \to \Int$ defined by
    \[
    (g \fconv  h)(\bv) \deff \sum_{\bu\in L^n,~\bw\in R^n} \iverson{
        \bv = \bu \oplus_f \bw} \cdot g(\bu) \cdot h(\bw)
    \]
    for every $\bv\in T^n$.
\end{definition}

As before the operations are taken in the standard $\Int(+,\cdot)$ ring
and $M$ is the maximum absolute value of the integers given on the input.

Now, we formally define the input and output to the \fConv problem.

\begin{definition}[\textsc{$f$-Convolution Problem (\fConv)}]
    Let $L$, $R$ and $T$ be arbitrary finite sets and let $f\from L \times R \to T$ be an arbitrary function.
  The \textnormal{\textsc{$f$-Convolution Problem}} is the following.
	\\
	\textbf{Input:}
	Two functions $g\from R^n\to \Mrange$ and $h\from L^n\to \Mrange$.
	\\
	\textbf{Task:}
	Compute $g \fconv h$.
\end{definition}

Our main result stated in the most general form is the following.

\begin{theorem}
    \label{thm:general-main}
  Let $f\from L\times R\to T$ such that  $L$, $R$ and $T$ are finite.
  There is an algorithm for the \fConv problem with $\Oc{c^n}$ time, where
  \[
      c \deff
    \begin{cases}
        \frac{|L|}{2} \cdot \left(\abs{R} + \frac{|T|}{2}\right)
        & \text{if } \abs{L} \text{ is even} \\
        \frac{|L|-1}{2}  \cdot \left(\abs{R} + \frac{|T|}{2}\right) +|R|
        & \text{otherwise.}
    \end{cases}
  \]
\end{theorem}

Theorem~\ref{thm:main-thm} is a corollary of
Theorem~\ref{thm:general-main} by setting $L = R = T = D$.

The proof of \cref{thm:general-main} utilizes the notion of \emph{cyclic partition}.
For any $k\in \Nat$, let $\Int_k=\{0,1,\ldots, k-1\}$.  We say a function $f\from
A\times B\to C$ is {\em $k$-cyclic} if, up to a relabeling of the sets $A$, $B$
and $C$, it is an addition modulo $k$.  Formally, $f\from A\times B\to C$ is {\em
$k$-cyclic} if there are $\sigma_A\from A\to\Int_k$, $\sigma_B\from B\to\Int_k$, and
$\sigma_C \from \Int_k \to C$ such that
\[
  \forall a\in A, ~b\in B:~~~f(a,b) =
    \sigma_C\left( \sigma_A(a)+ \sigma_B(b)\mod k\right)
  .
\]
We refer to the functions $\sigma_A$, $\sigma_B$ and $\sigma_C$ as the {\em relabeling} functions of $f$.

The {\em restriction} of $f\from L\times R \to T$ to $A\subseteq L$ and $B\subseteq
R$ is the function $g\from A\times B\to T$ defined by $g(a,b) = f(a,b)$ for all $a\in
A$ and $b\in B$.  We say $(A,B,k)$ is a {\em cyclic minor} of $f\from L\times R \to
T$ if the restriction of $f$ to $A$ and $B$ is a $k$-cyclic function.

A {\em cyclic partition} of $f\from L\times R \to T$ is a set of minors
$\Part=\{(A_1,B_1,k_1),\ldots, (A_m,B_m,k_m)\}$ such that $(A_i,B_i,k_i)$ is a
cyclic minor of $f$ and for every $(a,b)\in L\times R$ there is a unique  $1\leq
i\leq m$ such that $(a,b)\in A_i\times B_i$.  The cost of the cyclic partition
is $\cost(\Part)=\sum_{i=1}^{m} k_i$.

\cref{thm:general-main} follows from the following lemmas.

\begin{restatable*}[Algorithm for Generalized Convolution]{lemma}{lemalgoconv}
	\label{lem:fastConvolutionUsingMinor}
  Let $L$, $R$ and $T$ be finite sets.
  Also, let $f\from L\times R\to T$ be a function
  and let $\Part$ be a cyclic partition of $f$.
  Then there is an $\Oc{(\cost(\Part)^n + |L|^n + |R|^n + |T|^n)}$ time algorithm for \fConv.
\end{restatable*}

\begin{restatable*}{lemma}{lemexistanceminor}
	\label{lem:existenceOfMinor}
	Let $f\from L\times R\to T$ where $L$, $R$ and $T$ are finite sets.
	Then there is a cyclic partition $\Part$ of $f$ such that
	$\cost(\Part)\leq \frac{\abs{L}}{2} \cdot (\abs{R} + \frac{\abs{T}}2)$ when $
    \abs{L}$ is even, and $\cost(\Part) \leq \abs{R} + \frac{\abs{L}-1}{2} \cdot (\abs{R} + \frac{\abs{T}}{2})$ when $\abs{L}$ is odd.
\end{restatable*}
The proof of~\cref{lem:fastConvolutionUsingMinor} is included in Section~\ref{sec:algo} and proof of~\cref{lem:existenceOfMinor}
is included in Section~\ref{sec:existenceMinor}.
The proof of \cref{lem:fastConvolutionUsingMinor} uses an algorithm for \CyConv.

\begin{definition}[\CyConv]\label{def:cyclic_conv}
Let  $k\in \mathbb{N}$ and  $\br \in \nat^k$. Also, let 
 $g,h\from Z\to \mathbb{N}$ be two  functions where $Z=\Int_{\br_1}\times\ldots
\times \Int_{\br_k}$.  The \CyConv of $g$ and $h$ is the function $(g\conv h):Z
\to \mathbb{N}$ defined by
\[
(g\conv h)(\bv)
= \sum_{\bu,\bw\in Z}
\left(\prod_{i=1}^k \iverson{\bu_i + \bw_i = \bv_i \mod
	\br_i}\right)\cdot g(\bu)\cdot h(\bw)
\]
for every $\bv \in Z$. 
\end{definition}

For any $K\subseteq \mathbb{N}$ we define the $K$-$\CyConvProb$ in which we restrict the entries of the vector $\br$ in \Cref{def:cyclic_conv} to be in $K$.  
\begin{definition}[$K$-$\CyConvProb$]
For any $K\subseteq \mathbb{N}$ the $K$-\CyConvProb is defined as follows.
\\
\textbf{Input:}
Integers $k, M \in \mathbb{N}$, a 
vector $\br \in \nat^k$ such that $\br_j \in K$ for every $j\in [k]$  and  two 
functions $g,h\from Z\to\Mrange$ where $Z=\Int_{\br_1}\times\ldots
\times \Int_{\br_k}$.
\\
\textbf{Task:}
Compute  the $\CyConv$  $g\conv h \from Z \to\Int$.
\end{definition}
Van Rooij \cite{Rooij20} claimed that the $\mathbb{N}$-\CyConvProb can be solved in
$\Occ{\left(\prod_{i=1}^k \br_i\right)}$ time.  However, for his algorithm to
work it must be given an appropriate large prime $p$ and several primitives
roots of unity in $\FF_p$. We are unaware of a method which deterministically
finds such a prime and roots  while retaining the running time.  To overcome
this obstacle we present an algorithm for the $K$-\CyConvProb
when $K\subseteq \mathbb{N}$ is a fixed finite set.
Our solution uses multiple smaller primes and the
Chinese Reminder Theorem. We include the details in
Appendix~\ref{sec:cyclic-convolution}.

\begin{theorem}[$K$-\CyConv]
	\label{thm:cyclic}
	For any finite set $K\subseteq \mathbb{N}$,
    there is an $\Occ{(\prod_{i=1}^k \br_i)}$ algorithm for the
    $K$-\CyConvProb.
\end{theorem}

\section{Generalized Convolution}
\label{sec:algo}

In this section we prove \cref{lem:fastConvolutionUsingMinor}.

\lemalgoconv

Throughout the section we fix $L$, $R$ and $T$, and $f \from L \times R \to T$ to be  as in the statement of
\cref{lem:fastConvolutionUsingMinor}.
Additionally, fix a cyclic partition $\Part = \{(A_1, B_1, k_1), \ldots, (A_m, B_m,
k_m)\}$. Furthermore, let $\sigma_{A,i}$, $\sigma_{B,i}$ and $\sigma_{C,i}$ be the
relabeling functions of the cyclic minor $(A_i,B_i,k_i)$ for every $i \in [m]$.
We assume the labeling functions are also fixed throughout this section.

In order to describe our algorithm for \cref{lem:fastConvolutionUsingMinor}, we
first need to establish several technical definitions.

\begin{definition}[Type]
    The {\em type} of two vectors $\bu\in L^{n}$ and $\bw\in R^{n}$ is the
    unique vector $\type \in [m]^{n}$ for which $\bu_i \in A_{\type_i}$ and
    $\bw_i \in B_{\type_i}$ for all $i \in [n]$.   
\end{definition}
Observe that the type of two vectors is well defined as $\Part$ is a cyclic partition. 
For any type $\type \in \{1, \ldots, m\}^{n}$ we
define 
\begin{align*}
\domL \deff A_{\type_1} \times \dots \times
A_{\type_n}, && \domR \deff B_{\type_1} \times \dots \times
B_{\type_n}, 
    && \domZ \deff \Int_{k_{\type_1}} \times \dots \times
    \Int_{k_{\type_n}}
\end{align*}
to be vector domains restricted to type $\type$. For any type $\type$ 
we  introduce  relabeling  functions on its  restricted domains.
The relabeling functions of $\type$ are the functions
$\sigmaL \from \domL \to \domZ$,
$\sigmaR \from \domR \to \domZ$, and
$\sigmaT \from \domZ \to T^{n}$
 defined as follows:%
\begin{align*}
  \sigmaL (\bv) & \deff \left( \sigma_{A, \type_1}(\bv_1), \ldots,
  \sigma_{A, \type_n}(\bv_n) \right)  &
  \forall \bv \in \domL,\\
  \sigmaR (\bv) & \deff \left( \sigma_{B, \type_1}(\bv_1), \ldots,
  \sigma_{B, \type_n}(\bv_n) \right) &
  \forall \bv \in \domR,\\
  \sigmaT (\bq) & \deff \left( \sigma_{C, \type_1}(\bq_1), \ldots,
  \sigma_{C, \type_n}(\bq_n) \right)  &
  \forall \bq \in \domZ.
\end{align*}
Our algorithm heavily depends on constructing the following projections.
\begin{definition}[Projection of function]
    The projection of a function 
     $g\from L^{n} \to \Int$  with respect to the type $\type \in [m]^{n}$, is the
     function $g_\type \from \domZ
    \to \Int$ defined as
    \begin{align*}
          g_{\type}(\bq) \deff \sum\limits_{\bu \in \domL}\iverson{ \sigmaL(\bu) = \bq} \cdot g(\bu)  
          && \text{ for every } \bq \in \domZ.
    \end{align*}
    Similarly, the projection $h_\type \from \domZ \to \Int$
		of a function $h \from R^{n} \to \Int$
		with respect to the type $\type \in [m]^n$ is defined as
    \begin{align*}
          h_{\type}(\bq) \deff \sum\limits_{\bw \in \domR}\iverson{ \sigmaR(\bw) = \bq} \cdot h(\bw)  
          && \text{ for every } \bq \in \domZ.
    \end{align*}
\end{definition}
The projections are useful due to the following connection with $g\fconv h$.  
\begin{lemma} \label{lem:equal}
	Let $g \from L^{n} \to \Int$ and $h \from R^{n} \to \Int$, then for every
	$\bv \in T^n$ it holds that:
	\begin{align*}
		\left( g \fconv h\right)(\bv) = \sum_{\type \in [m]^{n}}
		\sum_{\bq\in \domZ} \iverson{\sigmaT(\bq) = \bv} \cdot \left( g_\type
		\conv h_\type \right)(\bq) 
		,
	\end{align*}
	where $g_\type \conv h_\type$ is the cyclic convolution of $g_\type$ and $h_\type$.  
\end{lemma}
We give the proof of \cref{lem:equal} in \cref{sec:projection}.
It should be noted that the naive computation
of the projection functions of $g$ and $h$ with respect to all types $\type$
is significantly slower than the running time stated in \cref{lem:fastConvolutionUsingMinor}.
To adhere to the stated running time
we use a dynamic programming procedure for the computations,
as stated in the following lemma.
\begin{lemma}\label{lemma:projection}
	There exists an algorithm which given a function $g\from L^n \to \Mrange$
	returns the set of its projections, $\{g_\type \mid \type \in [m]^{n} \}$, in time
    $\Occ{\left(\cost(\Part)^n + |L|^n\right)}$. 
\end{lemma}
\begin{remark}
	Analogously, we can also construct every projection of a function $h \from R^n \to
    \Mrange$ in $\Occ{\left(\cost(\Part)^n + |R|^n\right)}$ time.
\end{remark}
The proof of \cref{lemma:projection} in given in \cref{sec:projection}.

Our algorithm for $\fConv$ (see~\cref{algo:using_minor} for the
pseudocode) is a direct implication of \cref{lem:equal} and \cref{lemma:projection}.
First, the algorithm computes the projections of $g$ and $h$  with respect to
every type $\type$.  Subsequently, the cyclic convolution of $g_\type$ and
$h_\type$ is computed efficiently  as described in \cref{thm:cyclic}. Finally,
the values of $(g \fconv h)$ are reconstructed by the formula in Lemma~\ref{lem:equal}.

\begin{algorithm}
	\DontPrintSemicolon
	\SetKwInOut{Setting}{Setting}
	\SetKwComment{Comment}{$\triangleright$\ }{}
	\Setting{Finite sets $L$, $R$ and $T$, $f \from L \times R \to T$ and a cyclic partition $\Part$ of $f$, of size~$m$.}
	\KwIn{$g\from L^n\to \Mrange$, $h\from R^n\to \Mrange$}
	
	Construct the projections of $g$ and $h$ w.r.t $\type$, for all $\type \in
	[m]^{n}$\label{algoline:projection}\Comment*[r]{\cref{lemma:projection} }
	
	For every $\type\in [m]^n$ compute $\mathsf{c}_\type =  g_\type \conv  h_\type $ \label{algoline:convs}\Comment*[r]{Cyclic convolutions (\cref{def:cyclic_conv})}
	
	Define $\mathsf{r}\from T^n\to \Int$ by \label{algoline:compute}
	\[
	\mathsf{r}(\bv) =\sum_{\type \in [m]^{n}}
	\sum_{~~\bq \in \domZ  \text{ s.t. } \sigmaT(\bq)
		= \bv~~} \mathsf{c}_{\type}(\bq)~~~~~~~~\textnormal{ for all $\bv\in T^n$}.
	\]

	\Return{$\mathsf{r}$}\;
	\caption{Cyclic Partition Algorithm for the $\fConv$ problem}
	\label{algo:using_minor}
\end{algorithm}

\begin{proof}[Proof of Lemma~\ref{lem:fastConvolutionUsingMinor}]
	
	Observe that Algorithm~\ref{algo:using_minor} returns  $\mathsf{r} \from T^n\to \Int$ such that for every $\bv\in T^n$ it holds that 
	\begin{align*}
		\mathsf{r}(\bv) = 
		\sum_{\type \in [m]^{n}}
		\sum_{\substack{\bq \in \domZ \\ \text{s.t. } \sigmaT(\bq)
				= \bv}} \mathsf{c}_{\type}(\bq)
		=
		\sum_{\type \in [m]^{n}} \sum_{\bq \in \domZ}
		\iverson{\sigmaT(\bq) = \bv} \cdot \left( g_\type \conv h_\type
		\right)(\bq)=\left( g \fconv h\right)(\bv),
	\end{align*}
	where the last equality is by \cref{lem:equal}.
    Thus, the algorithm returns $(g\fconv h)$ as required. It therefore remains
    to bound the running time of the algorithm.
	
	By Lemma~\ref{lemma:projection},  Line~\ref{algoline:projection} of
    Algorithm~\ref{algo:using_minor} runs in time $\Oc{(\cost(\Part)^n +
    |L|^n+|R|^n)}$. Define  $K=\{ k \mid (A,B,k) \in \Part \}$ be different costs of cyclic minors in $\Part$. 
	By \cref{thm:cyclic}, for any type $\type \in[m]^n$ the computation of
    $g_\type \conv h_{\type}$ in Line~\ref{algoline:convs} is an instance of $K$-\CyConvProb which can be solved in time
$\Oc{(\prod_{ i=1}^{n} k_{\type_i}) } $. Thus the overall running time of
    Line~\ref{algoline:convs} is $\Occ{ (\sum_{\type \in [m]^{n} } \prod_{i=1}^n
    k_{\type_i}) }$.
	
	Finally, observe that the construction of $\mathsf{r}$ in Line~\ref{algoline:compute}  
	can be implemented by initializing $\mathsf{r}$ to be zeros and iteratively
    adding the value of $\mathsf{c}_{\type}(\bq)$ to
    $\mathsf{r}(\sigma^T_{\type}(\bq))$ for every $\type\in [m]^n$ and $\bq\in
    \domZ$. The required running time is thus $\Oc{|T|^n}$ for the
    initialization and $\Occ{(\sum_{\type\in [m]^n} |\domZ|)}=\Occ{(\sum_{\type\in
    [m]^n} \prod_{i=1}^{n} k_{\type_i})}$ for the addition operations. Thus, the overall running time of Line~\ref{algoline:compute} is 
    \begin{displaymath}
      \Occ{\left( |T|^n+ \sum_{\type \in [m]^{n} } \prod_{i=1}^n k_{\type_i}\right) }.  
    \end{displaymath}
	Combining the above,  
    with $\sum_{\type \in [m]^{n} }
    \prod_{i=1}^n k_{\type_i}= \left(\sum_{i=1}^{m} k_i \right)^n =
    \left(\cost(\Part)\right)^n$ means that
    the running time of Algorithm~\ref{algo:using_minor} is 
    \begin{displaymath}
        \Occ{\left(|T|^n + |R|^n + |L|^n+\cost(\Part)^n\right)}
    \end{displaymath}

	This concludes the proof of~\cref{lem:fastConvolutionUsingMinor}.
\end{proof}

\subsection{Properties of Projections}
\label{sec:projection}
In this section we provide the proofs for \cref{lem:equal} and \cref{lemma:projection}.
The proof of \cref{lem:equal} uses the following definitions of coordinate-wise  addition with respect to a type $\type$.

\begin{definition}[Coordinate-wise addition modulo for type]
	For any $\type \in [m]^n$ we define a coordinate-wise
	addition modulo as 
	\begin{align*}
		\bq +_\type \br \deff \left( (\bq_1 + \br_1 \mod
		k_{\type_1}),\ldots,(\bq_n + \br_n \mod k_{\type_n}) \right) &&
		\text{ for every } \bq,\br \in \domZ.
	\end{align*}
	
\end{definition}

\begin{proof}[Proof of \cref{lem:equal}]
	By Definition~\ref{def:fconv} it holds that:
	\begin{equation}\label{eq:basicdef}
		\left( g \fconv h\right)(\bv) = \sum_{\bu \in L^{n}, \bw \in R^{n}}
        \iverson{\bv = \bu \oplus_f \bw} \cdot g(\bu) \cdot h(\bw).
\end{equation}
	Recall that the type of every two vectors $(\bu,\bw)\in L^n \times R^n$ is unique
	and $[m]^n$ contains all possible types and
    hence, we can rewrite \eqref{eq:basicdef} as
	\begin{align}
        \label{eq:rewrite_basic}
		( g \fconv h)(\bv) &=
		\sum_{\type \in [m]^n} \sum_{\bu\in \domL, \bw\in \domR
				% \textnormal{s.t.   the type of $\bu$ and $\bw$ is $\type$
		}
			g(\bu)\cdot h(\bw) \cdot \iverson{\bv= \bu \oplus_f\bw }
		\intertext{By the properties of the relabeling functions, we get}
		\nonumber
		 &=\sum_{\type \in [m]^{n}} \sum_{\bu \in
		\domL, \bw \in \domR} g(\bu) \cdot h(\bw) \cdot \iverson{\bv =
		\sigmaT\left( \sigmaL(\bu) +_\type \sigmaR(\bw) \right) }
			\\\nonumber 
            &= \sum_{\type \in [m]^{n}} \sum_{\bq \in \domZ}
			\sum_{\bu \in \domL, \bw \in \domR} g(\bu) \cdot h(\bw) \cdot
			\iverson{\bv = \sigmaT(\bq)}\cdot \iverson{\bq = \sigmaL(\bu) +_\type
			\sigmaR(\bw)}\\ \nonumber
			&= \sum_{\type \in [m]^{n}}
			\sum_{\substack{\bq \in \domZ \\ \text{ s.t. } \sigmaT(\bq)
			= \bv}} \sum_{~\bu \in \domL, \bw \in \domR~} g(\bu) \cdot h(\bw)
			\cdot \iverson{\bq = \sigmaL(\bu) +_\type
            \sigmaR(\bw)}.
	\end{align}
    Observe that we can \emph{partition} $\domL$ (respectively $\domR$) by
    considering the inverse images of $\br \in \domZ$ under $\sigmaL$ (respectively
    $\sigmaR$), i.e. $\domL = \biguplus_{\br \in \domZ} \{\bu \in \domL \mid \sigmaL(\bu) = \br\}$. Hence, for every $\type \in [m]^n$  and $\bq \in \domZ$ it holds that
    \begin{align}
    	\label{eq:second_rewrite}
        \nonumber
	 &\sum_{~\bu \in \domL, \bv \in \domR~} g(\bu) \cdot h(\bw)
	\cdot \iverson{\bq = \sigmaL(\bu) +_\type
		\sigmaR(\bw)}\\\nonumber
	=& \sum_{\br, \bs \in \domZ}  \sum_{~\bu \in \domL, \bw \in \domR~} g(\bu) \cdot h(\bw)
	\cdot \iverson{\bq = \br +_\type
		\bs}\cdot  \iverson{\br = \sigmaL(\bu)} \cdot \iverson{\bs=\sigmaR(\bw)}\\\nonumber
	=&\sum_{\br, \bs \in \domZ} \iverson{\bq = \br +_\type
		\bs} \left( \sum_{\bu \in \domL}\iverson{\br = \sigmaL(\bu)}\cdot g(\bu) \right)\cdot \left( \sum_{\bw \in \domR}   
 \iverson{\bs=\sigmaR(\bw)}\cdot h(\bw)\right) \\\nonumber
 =&\sum_{\br, \bs \in \domZ} \iverson{\bq = \br +_\type
 	\bs}\cdot  g_{\type}(\br)\cdot h_{\type} (\bs)\\
 =& \;(g_{\type} \conv h_{\type})(\bq).
    \end{align}
    
    By plugging \eqref{eq:second_rewrite} into \eqref{eq:rewrite_basic} we get 
	\begin{align*}
		\left( g \fconv h\right)(\bv) &= \sum_{\type \in [m]^{n}}
		\sum_{\substack{\bq \in \domZ \\ \text{ s.t. } \sigmaT(\bq)
				= \bv}} (g_{\type}\conv h_{\type})(\bq)= \sum_{\type \in [m]^{n}}
			\sum_{\bq\in \domZ} \iverson{\sigmaT(\bq) = \bv} \cdot \left( g_\type
		\conv h_\type \right)(\bq),
	\end{align*}
    as required.
\end{proof}

\begin{proof}[Proof of \cref{lemma:projection}]
	The idea is to use a dynamic programming algorithm loosely inspired by Yates
	algorithm~\cite{yates1937design}.

     Define $X^{(\ell)} =\left\{ (\type, \bq)~\middle|~\type \in [m]^{\ell},~\bq \in \Int_{\type_1} \times \dots \times \Int_{\type_\ell}\right\}$
     for every $\ell \in \{0,\ldots, n\}$.
 We use $X^{(\ell)}$ to 
  define a dynamic programming table
	$\DP^{(\ell)}\from
	X^{(\ell)} \times L^{n-\ell} \to \Int$ for every $\ell \in \{0,\ldots n\}$ by:
	\begin{align*}
		\DP^{(\ell)}[(\type_1, \ldots,
		\type_\ell),(\bq_1,\ldots,\bq_\ell)][\bt_{\ell+1},\ldots,\bt_{n}] \deff
		\sum_{\substack{\bt_1 \in A_{\type_1}\\\ldots\\\bt_\ell \in A_{\type_\ell}}}
		\left(\prod_{i=1}^\ell \iverson{\sigma_{\type_i}(\bt_i) = \bq_i} \right) \cdot 
		g(\bt_1, \ldots, \bt_n)
		.
	\end{align*}
	
	The tables $\DP^{(0)},\DP^{(1)},\ldots, \DP^{(n)}$ are computed
    consecutively where the computation of $\DP^{(\ell)}$ relies on the values of
    $\DP^{(\ell-1)}$ for any $\ell \in [n]$.
	Observe that
	$g_\type(\bq) =
	\DP^{(n)}[(\type_1,\ldots,\type_n),(\bq_1,\ldots,\bq_n)][\varepsilon]$ for every
	$\type$ and $\bq$, which
	means that computing $\DP^{(n)}$ is equivalent to computing the projection functions $g_{\type}$ of $g$ for every type $\type$.\footnote{We use $\varepsilon$
		to denote the vector of length $0$.}

	It holds that $\DP^{(0)}[\varepsilon,\varepsilon][\bt] = g(\bt)$. Hence,
	$\DP^{(0)}$ can be trivially computed in $|L|^n$ time.  We use the following
	straightforward recurrence to compute $\DP^{(\ell)}$:
	\begin{equation}
		\label{eq:relation}
        \begin{aligned}
		&\DP^{(\ell)} [(\type_1, \ldots,
		\type_\ell),(\bq_1,\ldots,\bq_\ell)][\bt_{\ell+1},\ldots,\bt_{n}]
		= \\%\nonumber
		&~~~~~~~~~~~~~~~~~\sum_{\bt_\ell \in A_{\type_\ell}}
		\iverson{\sigma_{\type_\ell}(\bt_\ell) = \bq_\ell} \cdot 
		\DP^{(\ell - 1)}[(\type_1, \ldots,
        \type_{\ell-1}),(\bq_1,\ldots,\bq_{\ell-1})][\bt_{\ell},\ldots,\bt_{n}].
        \end{aligned}
	\end{equation}
A dynamic programming algorithm  which computes $\DP^{(n)}$ can be easily derived from~\eqref{eq:relation} and the formula for $\DP^{(0)}$.   The total number of states in the
	dynamic programming table $\DP^{(\ell)}$ is 
	\begin{align*}
		\left(\sum_{\type \in [m]^{\ell}} \left( k_{\type_1} \cdot \ldots\cdot
		k_{\type_\ell}\right) \right) \cdot \abs{L}^{n-\ell} = \left( k_1 + \dots + k_m
		\right)^{\ell} \cdot \abs{L}^{n-\ell} &= \cost(\Part)^{\ell} \cdot
		\abs{L}^{n-\ell}.
	\end{align*}
	This is bounded by $\cost(\Part)^n + |L|^n$ for every $\ell \in [n]$.
	To transition between states we spend polynomial time per entry because we
	assume that $|L| = \Oh(1)$. Hence, we
    can compute $g_\type$ for every $\type$ in $\Oc{(\cost(\Part)^n + |L|^n)}$ time.
\end{proof}

\section{Existence of Low-Cost Cyclic Partition}
\label{sec:existenceMinor}

In this section we prove \cref{lem:existenceOfMinor}.

\lemexistanceminor

We first consider the special case when $\abs{L}=2$.  Later we reduce the general case
to this scenario and use the result as a black-box.

As a warm-up we construct a cyclic partition of cost at most $\frac 7 8
\abs{\Dd}^2$ assuming that $L=R=T=\Dd$ and that $\abs{\Dd}$ is even.  For this,
we first partition $\Dd$ into pairs $d_1^{(i)},d_2^{(i)}$ where
$i\in[\abs{\Dd}/2]$ and show for each such pair that $f$ restricted to
$\{d_1^{(i)},d_2^{(i)}\}$ and $\Dd$ has a cyclic partition of cost at most
$\frac 7 4 \abs{\Dd}$.  The union of these cyclic partitions forms a cyclic
partition of $f$ with cost at most $\frac{\abs{\Dd}}2 \cdot \frac 7 4 \abs{\Dd}
= \frac{7}{8} \abs{\Dd}^2$.

To construct the cyclic partition for a fixed $i\in[\abs{\Dd}/2]$,
we find a maximal number $r$ of pairwise disjoint pairs
$e_1^{(j)},e_2^{(j)} \in \Dd$
such that $\abs{\{f(d_{a}^{(i)},e_b^{(j)}) \mid a,b \in \{1,2\} \}}\le 3$
for each $j \in [r]$,
i.e.\ for each $j$ at least one of the four values
$f(d_1^{(i)},e_1^{(j)}),f(d_1^{(i)},e_2^{(j)}),f(d_2^{(i)},e_1^{(j)}),f(d_2^{(i)},e_2^{(j)})$ repeats.
With this assumption,
$f$ restricted to $\{d_1^{(i)},d_2^{(i)}\}$ and $\{e_1^{(j)},e_2^{(j)}\}$
is either a cyclic minor of cost at most $3$
or can be decomposed into $3$ trivial cyclic minors of the total cost at most $3$.
We claim that $r \ge \abs{\Dd}/4$.
Indeed, assume that there are fewer than $|D|/4$ such pairs,
i.e.\ $r < \abs{\Dd}/4$.
Let $\overline D$ denote the $\abs{\Dd}-2 \cdot r > \abs{\Dd}/2$ remaining values in $\Dd$.
As the set $\{f(d_a^{(i)}, d) \mid d \in \overline \Dd, a \in \{1,2\} \}$
can only contain at most $\abs{\Dd}$ values,
we can find another pair $e_1^{(r+1)},e_2^{(r+1)}$ with the above constraints.
Note that $f$ restricted to $\{d_1^{(i)},d_2^{(i)}\}$ and $\overline D$
can be decomposed into at most $2\abs{\overline \Dd}$ trivial minors.
Hence,
the cyclic partition for $f$ restricted to $\{d_1^{(i)},d_2^{(i)}\}$ and $\Dd$
has cost at most
\[
  3 r + 2 \cdot \abs{\overline \Dd}
  \le 3 \cdot \frac{\abs{\Dd}}4 + 2 \cdot \frac{\abs{\Dd}}2
  \le \frac 7 4 \abs{\Dd}
  .
\]

\subsection{Special Case: $|L| = 2$}

In this section, we prove the following lemma that is a special case
of~\cref{lem:existenceOfMinor}.

  \begin{lemma}
    \label{lem:existenceMinor:twoRows}
    If $f\from L \times R \to T$ with $\abs{L}=2$,
    then there is a cyclic partition $\Part$ of $f$
    such that $\cost(\Part) \le \abs{R} + \abs{T}/2$.
  \end{lemma}

To construct the cyclic partition we proceed as follows. First, we define,
for a function $f$, the representation graph $G_f$.  Next, we show that if
this graph has a special structure, which we later call \emph{nice}, then we
can easily find a cyclic partition for the function $f$.  Afterwards we
decompose (the edges of) an arbitrary representation graph $G_f$ into nice
structures and then combine the cyclic partitions coming from these parts to a
cyclic partition for the original function $f$.

\begin{definition}[Graph Representation]
  Let $f\from L \times R \to T$ be such that $\abs{L}=2$
  with $L = \{\ell_0, \ell_1\}$.

  We say a function $\lambda_f\from R \to T \times T$
  with $\lambda_f\from r \mapsto (f(\ell_0,r),f(\ell_1,r))$
  is the \emph{edge mapping} of $f$.
  We say that a directed graph $G_f$ (which might have self-loops)
  with vertex set $V(G_f) \deff T$
  and edge set $E(G_f) \deff \{ \lambda_f(r) \mid r \in R\}$
  is the \emph{representation graph} of $f$.

  We say that the representation graph $G_f$ is \emph{nice} if $G_f$ is a
  directed cycle or a directed path (potentially with a single
  edge).
\end{definition}

\begin{definition}[Restriction of $f$]
  Let $f\from L \times R \to T$ be a such that $\abs{L}=2$
  % with $L = \{ \ell_0, \ell_1 \}$
  and let $G_f$ be a graph representation of $f$.

  Let $E'\subseteq E(G_f)$ be a subset of edges
  inducing the subgraph $G'$ of $G_f$.
  With $T'\deff V(G')$ and
  $R'\deff \{ r \in R \mid \lambda_f(r) \in E'\}$,
  we define $f' \from L \times R' \to T'$ as the \emph{restriction} of $f$
  such that the representation graph of $f'$ is $G'$.
  Formally,
  $f'(\ell,r)=f(\ell,r)$ for all $\ell\in L$ and $r \in R'$.
  We say that $f'$ is the \emph{function represented by} $G'$ or $E'$, respectively.
\end{definition}

A decomposition of a directed graph $G$ is a family $\mathcal{F}$ of edge-disjoint
subgraphs of $G$, such that each edge belongs to exactly one subgraph in $\mathcal{F}$.
% For ease of notation,
% we identify the set of edges $E'$ with the induced graph $G'$. \karol{Do we really have $G'$ and $E'$ further in the text?}
The following observation follows directly from the previous definition.

\begin{observation}
  \label{obs:combinePartitions}
  Let $\{G_1,\dots,G_k\}$ be a decomposition of the graph $G_f$ into $k$ subgraphs,
  let $f_i$ be the function represented by $G_i$,
  and let $\Part_i$ be a cyclic partition of $f_i$.
  Then $\Part=\bigcup_{i\in [k]} \Part_i$ is a cyclic partition of $f$
  with cost $\cost(\Part) = \sum_{i\in [k]} \cost(\Part_i)$.
\end{observation}

  \begin{figure}[t]
    \centering
    \includegraphics[width=0.8\textwidth]{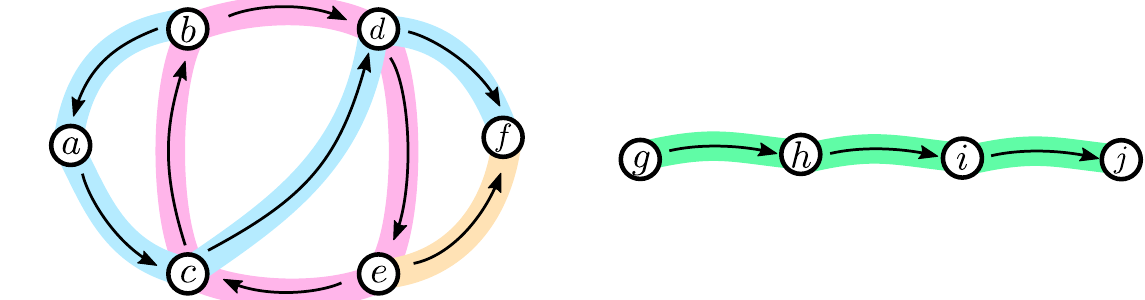}
    \centering

    \vspace*{0.1cm}
    \begin{tabular}{ccccccccccccc}
    & $r_1$                  & $r_2$
    & $r_3$                  & $r_4$
    & $r_5$                  & $r_6$
    & $r_7$                  & $r_8$
          & $r_9$                  & $r_{10}$
          & $r_{11}$                  & $r_{12}$
    \\ \cline{2-13}
    \multicolumn{1}{c|}{$\ell_0$} &
    \multicolumn{1}{c|}{$a$} &
    \multicolumn{1}{c|}{$b$} &
    \multicolumn{1}{c|}{$b$} &
    \multicolumn{1}{c|}{$c$} &
    \multicolumn{1}{c|}{$c$} &
    \multicolumn{1}{c|}{$d$} &
    \multicolumn{1}{c|}{$d$} &
    \multicolumn{1}{c|}{$e$} &
    \multicolumn{1}{c|}{$e$} &
    \multicolumn{1}{c|}{$g$} &
    \multicolumn{1}{c|}{$h$} &
    \multicolumn{1}{c|}{$i$} \\ \cline{2-13}
    \multicolumn{1}{c|}{$\ell_1$} &
    \multicolumn{1}{c|}{$c$} &
    \multicolumn{1}{c|}{$a$} &
    \multicolumn{1}{c|}{$d$} &
    \multicolumn{1}{c|}{$b$} &
    \multicolumn{1}{c|}{$d$} &
    \multicolumn{1}{c|}{$e$} &
    \multicolumn{1}{c|}{$f$} &
    \multicolumn{1}{c|}{$c$} &
    \multicolumn{1}{c|}{$f$} &
    \multicolumn{1}{c|}{$h$} &
    \multicolumn{1}{c|}{$i$} &
    \multicolumn{1}{c|}{$j$} \\ \cline{2-13}
    \end{tabular}
    % \captionlistentry[table]{A table beside a figure}
    % \captionsetup{labelformat=andtable}
    \caption{
    Example of the construction of a representation graph from the function $f$ to obtain a cyclic partition.
    We put an edge between vertices $u$ and $v$
    if there is an $r_i$ with $u = f(\ell_0,r_i)$ and $v = f(\ell_1,r_i)$.
    We highlight an example decomposition of the edges into a cycle with $4$ vertices (highlighted red) and three paths
    with $5$, $2$ and $4$ vertices (highlighted
    blue, yellow and green respectively). The cost of this cyclic partition is
$4 + 5 + 2 + 4  = 15$. }
    \label{fig:exampleGraphForMinorPartition}
  \end{figure}

  \subparagraph*{Cyclic Partitions Using Nice Representation Graphs.}
  As a next step, we show that functions admit cyclic partitions
  if the representation graph is nice.
  We extend these results to functions with arbitrary representation graphs
  by decomposing these graphs into nice subgraphs.
  Finally, we combine these results to obtain a cyclic partition for the original function $f$.

\begin{lemma}
  \label{clm:existenceMinor:cycle}
  \label{clm:existenceMinor:path}
  \label{clm:existenceMinor:star}

	Let $f\from L \times R \to T$ be a function such that $G_f$ is nice.
	Then $f$ has a cyclic partition of cost at most $\abs{T}=\abs{V(G_f)}$.
\end{lemma}
\begin{proof}
	By definition,
  a nice graph is either a cycle or a path.
	We handle each case separately in the following.
  Let $L = \{\ell_0, \ell_1\}$.
  \begin{description}
  \item[$G_f$ is a cycle.]
    We first define the relabeling functions of $f$
    to show that $f$ is $\abs{T}$-cyclic.

    For the elements in $L$, let $\sigma_L \from L \to \Int_2$
    with $\sigma_L(\ell_i) = i$.
    To define $\sigma_R$ and $\sigma_T$,
    fix an arbitrary $t_0 \in T$.
    Let $t_1,\dots,t_{\abs{T}}$ be the elements in $T$
    with $t_{\abs{T}}=t_0$
    such that, for all $j\in \Int_{\abs{T}}$, there is some $r_j\in R$
    with $\lambda_f(r_j) = (t_j, t_{j+1})$.%
    \footnote{
    Note that there might be multiple $r \in R$ with $\lambda_f(r)=(t_j,t_{j+1})$.
    }
    Note that these $r_i$ exist since $G_f$ is a cycle.
    Using this notation, we define $\sigma_T \from \Int_{\abs{T}} \to T$
    with $\sigma_T(j) = t_j$,
    for all $j \in \Int_{\abs{T}}$.
    For the elements in $R$ we define
    $\sigma_R\from R \to \Int_{\abs{R}}$
    with $\sigma_R(r) = j$ whenever $\lambda_f(r)=(t_j, t_{j+1})$ for some $j$.

    It is easy to check that $f$
    can be seen as addition modulo $\abs{T}$.
    Indeed, let $i \in \{0,1\}$ and $r \in R$
    with $\lambda_f(r) = (t_j, t_{j+1})$.
    Then we get
    \[
      \sigma_T( \sigma_L(\ell_i) + \sigma_R(r) \bmod \abs{T})
      = \sigma_T(i + j \bmod \abs{T})
        = t_{(i+j \bmod \abs{T})}
        = f(\ell_i, r_j) = f(\ell_i, r)
        .
    \]
    Thus, $f$ is $\abs{T}$-cyclic
    and $\{ (L, R, \abs{T}) \}$ is a cyclic partition of $f$.
    
  \item[$G_f$ is a path.]
  	Similarly to the previous case,
    $f$ can be represented as addition modulo $\abs{T}$.
    As the proof is essentially identical to the cyclic case,
    we omit the details here.
    \qedhere
  \end{description}
\end{proof}

In the next step, we decompose arbitrary graphs into nice subgraphs.
To present our decomposition we need to introduce the following notation related to the degree of vertices.

\begin{definition}[Sources, Sinks and Middle Vertices]
  Let $G=(V,E)$ be a directed graph.
  We denote by $\indeg(v)$ the \emph{in-degree} of~$v$,
  i.e., the number of edges terminating at~$v$,
  and by $\outdeg(v)$ the \emph{out-degree} of $v$,
  i.e., the number of edges starting at $v$.

  We partition $V$ into the three sets $\Vsrc(G)$, $\Vmid(G)$, and $\Vsnk(G)$
  defined as follows:
  \begin{itemize}
    \item 
    Set $\Vsrc(G)$ contains all \emph{source} vertices of $G$,
    that is, vertices with no incoming edges
    (i.e., $\indeg(v)=0$).
    This includes all isolated vertices.

    \item
    Set $\Vmid(G)$ contains all \emph{middle} vertices of $G$,
    that is vertices with incoming and outgoing edges
    (i.e., $\indeg(v),\outdeg(v)\ge 1$).

    \item
    Set $\Vsnk(G)$ contains the (remaining) \emph{sink} vertices of $G$,
    that is, vertices with incoming but no outgoing edges
    (i.e., $\indeg(v)\ge 1$ and $\outdeg(v)=0$).
  \end{itemize}
\end{definition}
We additionally introduce the notion of \emph{deficiency}
which we use in the following proofs.

\begin{definition}[Deficiency]
  Let $G=(V,E)$ be a directed graph.
  For all $v\in V$, we denote by
  \(
    \defi(v) \deff \max \{ \outdeg(v) - \indeg(v), 0 \}
  \)
  the \emph{deficiency} of $v$.

  We define $\Dall(G) \deff \sum_{v \in V} \defi(v)$
  as the \emph{total deficiency} of the graph $G$.  
\end{definition}
We omit the graph $G$ from the notation
if it is clear from the context.

We use the deficiency to decompose the acyclic graphs into paths.
\begin{lemma}
  \label{prop:existenceMinor:decomposePaths}
  Every directed graph $G$ can be decomposed into $\Dall(G)$ paths
  and an arbitrary number of cycles.
\end{lemma}
\begin{proof}
    We construct the decomposition $\mathcal{F}$ of $G$ as follows.
    In the first phase, we exhaustively
    find a directed cycle $C$ in $G$.
    We add cycle $C$ to the decomposition $\mathcal{F}$ and
    remove the edges of $C$ from $G$.  We continue the above
    procedure until graph $G$ becomes acyclic. Next, in the second phase we exhaustively find a
    directed maximum length path $P$ (note that $P$ may be a single edge). We add $P$ to the decomposition
	$\mathcal{F}$ and remove the edges of $P$ from $G$. We repeat the second phase until the graph $G$ becomes edgeless.

    This concludes the construction of decomposition $\mathcal{F}$. For correctness
    observe that the above procedure always terminates because in each step we
    decrease the number of edges of $G$. Moreover, at the end of the above
    procedure $\mathcal{F}$ is a decomposition of $G$ that consists only of paths and
    cycles. 

    We are left to show that the number of paths in $\mathcal{F}$ is exactly $\Dall(G)$.
    Note that deleting a cycle in $G$ does not change the value of
    $\Dall(G)$, hence the first phase of the procedure does not influence
    $\Dall(G)$ and we can assume that $G$ is acyclic.

    Next, we show that deleting a maximum length path from an acyclic graph
    decrements its deficiency by exactly $1$. This then conclude the proof,
    because in the second phase of the procedure the deficiency of $G$ decreases
    from $\Dall(G)$ down to $0$, which means that exactly $\Dall(G)$ maximum length
    paths were added to $\mathcal{F}$.

	Let $P$ be a maximum length, directed path in the acyclic graph $G$.  Let
	$s,t \in V(G)$ be the starting and terminating vertices of path $P$.  Path $P$
	must start at a vertex with a positive deficiency, because otherwise $P$ could
	have been extended at the start which would contradict the fact that $P$ is of maximum length.
	Similarly, since $P$ is of maximum length it must terminate in a sink vertex. Hence $\defi(s) > 0$ and $\defi(t) = 0$.
	Moreover, every vertex $v \in P \setminus \{s,t\}$ has exactly one incoming and
	one outgoing edge in $P$. Therefore, in the graph $G \setminus P$ the
	contribution to the total deficiency decreased only in the vertex $s$ and only
	by $1$. This means that $\Dall(G) = \Dall(G \setminus P) + 1$ which concludes
	the proof.
\end{proof}

Now we combine \cref{clm:existenceMinor:cycle,clm:existenceMinor:path,prop:existenceMinor:decomposePaths}
to show \cref{clm:existenceMinor:second}.
\begin{lemma}
  \label{clm:existenceMinor:second}
  Let $f\from L \times R \to T$ be a function with $\abs{L}=2$
  and let $G_f$ be the representation graph of $f$.
  Then, there exists a cyclic partition $\Part$ for $f$
  with $\cost(\Part) \le \abs{E(G_f)} + \Dall(G_f) $.
\end{lemma}
\begin{proof}
  First, use \cref{prop:existenceMinor:decomposePaths}
  to decompose the graph into cycles and $\Dall(G_f)$ paths.
  Then, for each of these paths and cycles,
  use \cref{clm:existenceMinor:path} to obtain the cyclic minor.
  By \cref{obs:combinePartitions}, these minors form a cyclic partition
  for the function represented by $G_f$.
  Let $\Part$ be the resulting cyclic partition.

  It remains to analyze the cost of the cyclic partition $\Part$.  By
  construction, each cyclic minor in $\Part$ corresponds to a path or a cycle
  (possibly of length $1$).  By~\cref{clm:existenceMinor:cycle} the cost of a
  path or a cycle is the number of vertices it contains.  Thus, for a path, the
  cost is equal to the number of edges plus one, and for a cycle the cost is
  equal to the number of edges.  Hence, the cost of $\Part$ is bounded by the
  number of edges of $G_f$ plus the number of paths in the decomposition. The
  latter is precisely~$\Dall(G_f)$
  by~\cref{prop:existenceMinor:decomposePaths}.
\end{proof}

\subparagraph*{Cyclic Partitions Using a Direct Construction.}
In the following, we use a different method
to construct a cyclic partition of the function $f$.
Instead of decomposing the graph into nice subgraphs,
we directly construct a partition and bound its cost.
\begin{lemma}
  \label{clm:existenceMinor:first}
  Let $f\from L \times R \to T$ be a function with $\abs{L}=2$
  and let $G_f$ be the representation graph of $f$.
  Then, there is a cyclic partition $\Part$ of $f$
  with $\cost(\Part) \le \abs{V(G_f)} + \abs{\Vmid(G_f)}$.
\end{lemma}
\begin{proof}
  For each $\ell \in L$, we use a single cyclic minor.
	Let $L=\{\ell_0, \ell_1\}$.
  For $i\in \{0,1\}$ define $T_i = \{f(\ell_i,r) \mid r\in R\}$
  and $k_i =\abs{T_i}$.
  Then, $\Part \deff \{ (\ell_i , R, k_i) \mid i\in \{0,1\}\}$
  is the cyclic partition of $f$.

  To see that $(\{\ell_i\}, R, k_i)$ is a cyclic minor for $i\in \{0,1\}$,
  assume w.l.o.g.\ that $T_i = \{0,1,\ldots, k_i-1\}$
  and define $\sigma_L(\ell_i)=0$, $\sigma_R(r) = f(\ell_i,r)$,
  and $\sigma_T(t) =t$.
  Thus, $\Part$
  is a cyclic partition of $f$ of cost $k_0+k_1 = |T_0|+|T_1|$.
	
  Observe that $\abs{T_0} = \abs{\Vsrc(G_f)} + \abs{\Vmid(G_f)}$
  as every $t\in T_0$ has an outgoing edge in $G_f$,
  and $\abs{T_1} = \abs{\Vsnk(G_f)} + \abs{\Vmid(G_f)}$
  as every $t\in T_1$ has an incoming edge in $G_f$.
  Hence,
  \begin{align*}
    \cost(\Part)
    &= \abs{T_0}+\abs{T_1}\\
    &= \abs{\Vsrc(G_f)}+\abs{\Vmid(G_f)}+ \abs{\Vsnk(G_f)} + \abs{\Vmid(G_f)}\\
    &= |V(G_f)| +\abs{\Vmid(G_f)}
  \end{align*}
  which finishes the proof.
\end{proof}

\subparagraph*{Bounding The Cost of Cyclic Partitions.}

Now, we combine the results from
\cref{clm:existenceMinor:first,clm:existenceMinor:second}.
We first show how the number of edges relates to the total deficiency of a graph and the number of middle vertices.

\begin{lemma}
  \label{prop:existenceMinor:structProp}
  For every directed graph $G$ it holds that \( \abs{\Vmid(G)} + \Dall(G) \le \abs{E(G)} \).
\end{lemma}
\begin{proof}
  Let $m$ be the number of edges of $G$ and let $e_1,\ldots,e_m \in E(G)$ be
  some arbitrarily fixed order of its edges. For every $i \in
  \{0,\ldots,m\}$ let $G_i$ be the graph with vertices $V(G)$ and edges
  $E(G_i) = \{e_1,\ldots,e_i\}$. Hence $G_0$ is an independent set of $V(G)$ and
  $G_m = G$.

  For every $i \in \{0,\ldots,m\}$ let $\Pot(G_i) \deff \abs{\Vmid(G_i)} + \Dall(G_i)$ be the quantity we need to bound.
  We show that
  \begin{equation}
      \label{eq:psi}
      \Pot(G_i) - \Pot(G_{i-1}) \le 1 \text{ for every } i \in [m]
  \end{equation}
  which then concludes the proof because
  \begin{displaymath}
      \abs{\Vmid(G)} + \Dall(G) = \Pot(G_m) = \sum_{i=1}^m
      \left(\Pot(G_i) - \Pot(G_{i-1})\right) \le m = \abs{E(G)}
      .
  \end{displaymath}
  From now, we focus on the proof of \cref{eq:psi}. For every $v \in V(G)$ and $i \in \{0,\ldots,m\}$, let $\defi_i(v)$ be the
  deficiency of vertex $v$ in graph $G_i$. Next, for every $v \in V(G)$ and $i
  \in [m]$, we define
  \begin{displaymath}
      \Delta_i(v) \deff  \defi_i(v) - \defi_{i-1}(v)  + \iverson{v \in \Vmid(G_i) \setminus \Vmid(G_{i-1})}
  \end{displaymath}
  Consider a step $i \in [m]$. Let $e_i = (s,t)$ be an $i$th edge that starts at
  a vertex $s$ and terminates at a vertex $t$. It holds that
  \begin{align*} 
      \abs{\Vmid(G_i)} + \Dall(G_i) = \abs{\Vmid(G_{i-1})} + \Dall(G_{i-1}) + \Delta_i(s) + \Delta_i(t).
  \end{align*}
  Therefore $\Pot(G_i) - \Pot(G_{i-1}) = \Delta_i(s) + \Delta_i(t)$ and to
  establish \cref{eq:psi} it is enough to show that $\Delta_i(s) \le 1$ and $\Delta_i(t) \le 0$.

  \begin{claim}
    \label{clm:structProp:start}
    It holds that $\Delta_i(s) \le 1$.
  \end{claim}
  \begin{claimproof}
    We consider two cases depending on whether $u$ became a middle vertex.
    If it happened that $s \in \Vmid(G_i) \setminus \Vmid(G_{i-1})$, then
    $s \in \Vsnk(G_{i-1})$ which means that
    $s$ has more incoming than outgoing edges in $G_{i-1}$. Hence
    $\defi_{i-1}(s) = \defi_i(s) = 0$ and we conclude that $\Delta_i(s)=1$.

    Otherwise $s \notin \Vmid(G_i)\setminus \Vmid(G_{i-1})$.
    Because the edge $e_i$ starts at $s$, 
    the deficiency of $s$ can increase by at most $1$.
    Hence, by $(\defi_i(s) - \defi_{i-1}(s)) \le 1$
    we conclude that $\Delta_i(s) \le 1$.
    \claimqedhere
  \end{claimproof}
  Finally, we consider the end vertex $t$ of the edge $e_i$.
  \begin{claim}
    \label{clm:structProp:end}
    It holds that $\Delta_i(t) \le 0$.
  \end{claim}
  \begin{claimproof}
    We again distinguish two cases
    depending on whether $t$ became a middle vertex.
    If $t \in \Vmid(G_i) \setminus \Vmid(G_{i-1})$, then $t \in \Vsrc(G_{i-1})$
    and moreover, $t$ has no incoming edges and the positive number of outgoing edges
    in $G_{i-1}$. Therefore $\defi_i(t) = \defi_{i-1}(t)-1$ which means that
    $\Delta_i(t) \le 0$.
      
    It remains to analyse the case when $t \notin \Vmid(G_i) \setminus \Vmid(G_{i-1})$.
    Since the edge $e_i$ ends at $t$, the deficiency of $t$ cannot increase and $\defi_i(v) \le \defi_{i-1}(v)$. This means that $\Delta_i(t) \le 0$.
      \claimqedhere
  \end{claimproof}
  By \cref{clm:structProp:start,clm:structProp:end},
  it follows that $\Delta_i(s) + \Delta_i(t) \le 1$.
  This establishes \cref{eq:psi} and concludes the proof.
\end{proof}
Now we are ready to combine
\cref{clm:existenceMinor:first,clm:existenceMinor:second}
and prove \cref{lem:existenceMinor:twoRows}.
\begin{proof}[Proof of \cref{lem:existenceMinor:twoRows}]

  As before, we denote by $G_f$ the representation graph of $f$. Let $V$ and
	$E$ be the set of vertices and edges of graph $G_f$.

  Let $\Part_1$ be the cyclic partition of $f$
  from \cref{clm:existenceMinor:second}
  with cost at most $\abs{E}+\Dall(G_f)$
  and let $\Part_2$ be the cyclic partition of $f$
  from \cref{clm:existenceMinor:first}
  with cost at most $\abs{V} + \abs{\Vmid(G_f)}$.
  We define $\Part$ as the minimum cost partition among $\Part_1$ and $\Part_2$.
  This implies that
  \begin{align*}
      \cost(\Part) &\le \min \{ \cost(\Part_1),\cost(\Part_2) \}
      \le  \frac{\cost(\Part_1) + \cost(\Part_2)}{2} \\
      &\le  \frac{\abs{E} + \abs{V} + \abs{\Vmid(G_f)} + \Dall(G_f)}{2}.
  \end{align*}
  Next, we use  the inequality $\abs{\Vmid(G_f)} + \Dall(G_f)\le \abs{E}$
      from \cref{prop:existenceMinor:structProp}, and get
  \begin{align*}
      \cost(\Part) &\le \abs{E} + \frac{\abs{V}}{2}.
  \end{align*}
  Since $\abs{E} \le \abs{R}$ and $\abs{V}=\abs{T}$ this concludes
  the proof.
\end{proof}

\subsection{General case: Proof of~\cref{lem:existenceOfMinor}}

Now we have everything ready to prove the main result of this section.

\begin{proof}[Proof of \cref{lem:existenceOfMinor}]
  We first handle the case when $\abs{L}$ is even.
  We partition $L$ into $\lambda=\abs{L}/2$ sets $L_1,\dots,L_\lambda$
  consisting of exactly two elements. % (except for possibly $L_\lambda$).
  We use \cref{lem:existenceMinor:twoRows} to find a cyclic partition $\Part_i$ for each
  $f_i\from L_i \times R \to T$.
  By definition of the cyclic partition,
  $\Part = \bigcup_{i\in [\lambda]} \Part_i$ is a cyclic partition for $f$, hence
  it remains to analyze the cost of $\Part$.

  Observe that for each $G_i$ we have that $\abs{V_i} \le \abs{T}$
  and $\abs{E_i} \le \abs{R}$.
  By the definition of the cost of the cyclic partition,
  we immediately get that
  \[
    \cost(\Part)
      \le \sum_{i=1}^\lambda \cost(\Part_i)
      \le \lambda \cdot \left( \abs{R} + \frac{\abs{T}}{2}\right)
    .
  \]
  If $\abs{L}$ is odd,
  then we remove one element $\ell$ from $L$ and let $L_0=\{\ell\}$.
  There is a trivial cyclic partition $\Part_0$ for $f_0\from L_0\times R \to T$
  of cost at most $\abs{R}$.
  Then we use the above procedure to find a cyclic partition $\Part'$
  for the restriction of $f$ to $L\setminus\{\ell\}$ and $R$.
  Hence, setting $\Part = \Part_0 \cup \Part'$ gives a cyclic partition for $f$
  with cost
  \[
    \cost(\Part) \le \cost(\Part_0) + \cost(\Part')
  \le \abs{R} + \floor{\frac{\abs{L}}{2}} \left( \abs{R} + \frac{\abs{T}}{2} \right)
    .
    \qedhere
  \]
\end{proof}
\begin{remark}
  If $\abs{L}$ and $\abs{R}$ are both even,
  one can easily achieve a cost of
  \[
    \min\left(
      \frac{L}2 \cdot \left( \abs{R} + \frac{\abs{T}}{2} \right),
      \frac{R}2 \cdot \left( \abs{L} + \frac{\abs{T}}{2} \right)
    \right)
    =
    \frac{\abs{L} \cdot \abs{R}}2 + \frac{\abs{T}}{4} \cdot\min(\abs{L},\abs{R})
  \]
  by swapping the role of $L$ and $R$
  and considering the function $f'\from R \times L \to T$
  with $f'(r,\ell)=f(\ell,r)$ for all $\ell \in L$ and $r \in R$.
\end{remark}

\subsection{Tight Example: Lower bound on~\cref{lem:existenceMinor:twoRows}}

To complement the previous results, we show that \cref{lem:existenceMinor:twoRows} is tight.  That is, there is a function
$f\from L\times R\to T$ with $\abs{L}=2$ such that no cyclic partition $\Part$
of $f$ has smaller cost, i.e., $\cost(\Part) < \abs{R} + \abs{T}/2$.  In
particular, this demonstrates that to improve the constant $c \deff 3/4$ in
\cref{thm:main-thm} new ideas are needed.

\begin{figure}[t]
\begin{center}
    \subfloat{
\vspace*{0.5cm}
\begin{tabular}{ccccc}
	& $r_1$                  & $r_2$
	& $r_3$                  & $r_4$
	\\ \cline{2-5}
	\multicolumn{1}{c|}{$\ell_0$} &
	\multicolumn{1}{c|}{$a$} &
	\multicolumn{1}{c|}{$b$} &
	\multicolumn{1}{c|}{$c$} &
	\multicolumn{1}{c|}{$a$} 
	\\ \cline{2-5}
	\multicolumn{1}{c|}{$\ell_1$} &
	\multicolumn{1}{c|}{$b$} &
	\multicolumn{1}{c|}{$c$} &
	\multicolumn{1}{c|}{$d$} &
		\multicolumn{1}{c|}{$d$} 
\\ \cline{2-5}
\end{tabular}
    }
    \subfloat{
\hspace{1cm}
       \includegraphics[width=0.15\textwidth]{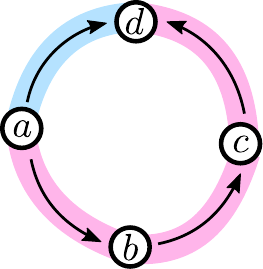}
    }
\end{center}
% \hfill
  \caption{
  The function $f$ from \cref{lem:existenceOfMinor:tightness}
  which shows that the bound from \cref{lem:existenceMinor:twoRows}
  is tight. The representation graph of $f$ is depicted on the right.
  We highlight
  the cyclic partition returned by \cref{lem:existenceOfMinor:tightness}.
  The red path contains
  $4$ vertices and the blue path contains $2$ vertices.
  Hence, the cost of that cyclic
  partition is $6$. \cref{lem:existenceOfMinor:tightness} shows that this is
  the best possible.}
  \label{fig:existenceOfMinor:tightness}
\end{figure}
\begin{lemma}
  \label{lem:existenceOfMinor:tightness}
  There exist sets $L$, $R$, and $T$ with $\abs{L}=2$
  and a function $f\from L\times R\to T$
  such that, every cyclic partition $\Part$ of $f$
  has $\cost(\Part) \ge \abs{R} + \abs{T}/2$.
\end{lemma}
\begin{proof}
  Define $L =\{\ell_0,\ell_1\}$,
  $R=\{r_1,r_2,r_3,r_4\}$, and $T=\{a,b,c,d\}$.
  Let $f$ be the function as defined in \cref{fig:existenceOfMinor:tightness}.
  Note that we need to show that every cyclic partition of $f$ has cost at least $6$.

  Let $\Part$ be a cyclic partition of $f$.  We first claim that the cyclic
  partition $\Part$ of $f$ contains a single cyclic minor,
  i.e., $\Part=\{(L,R,k)\}$ for some
  integer $k$.  For contradictions
  sake, we analyse every other remaining structure of $\Part$ and argue that in
  each case $\cost(\Part)\ge 6=\abs{R}+\abs{T}/2$.

  \begin{itemize}
  	\item
    Every cyclic minor in $\Part$ is of the form $(\{\ell_i\}, B, k)$
    (i.e., uses only values from a single row).
    Then, $\cost(\Part)\geq 6$ as each row has $3$ distinct values.

  	\item
    There is a cyclic minor $(\{\ell_0,\ell_1\}, \{r_j\}, k)$ in $\Part$.
    Since each column contains two distinct elements,
    It must hold that $k\geq 2$.
    Furthermore, the cyclic minors which cover the remainder of the graph
    must have a total cost of $4$ (or more)
    as all values in $T$ appear in the remainder of the graph.
    Hence $\cost(\Part)\geq 6$.

    \item
    There is a cyclic minor $(\{\ell_0, \ell_1\}, \{r_j,r_{j'}\},k)$ in $\Part$.
    Since each pair of two columns contains (at least) three values,
    it must hold that $k\geq 3$.
    There are at least $3$ distinct values in the remainder of the graph,
    hence, the cost of the remaining minors in $\Part$ is at least $3$.
    Thus $\cost(\Part)\geq 6$.

  	\item
    There is a cyclic minor $(\{\ell_0, \ell_1\}, R\setminus \{r_j\},k)$ in $\Part$.
    It holds that $k\geq 4$ as every three columns include all values in $T$.
    In each case, there are two different values in the remaining column.
    Hence, the cost of the remaining minors is at least $2$.
    Therefore~$\cost(\Part)\geq 6$.
  \end{itemize}

  With this, we know that $\Part$ contains only the single cyclic minor $(L,R,k)$.
  Let $\sigma_L, \sigma_R$ and $\sigma_T$ be the relabelling functions of $(L,R,k)$.
  From the definition of the relabeling functions, we get that
  $F\deff \{ \sigma_L(\ell_i) + \sigma_R(r_j) \mod k \mid i \in \{0,1\} \text{ and } j\in \{1,2,3\} \}$
  contains at least four elements.

  We claim that $(\sigma_L(\ell_0)+\sigma_R(r_4) \mod k) \notin F$.
  For the sake of contradiction assume otherwise.
  Then, by the definition of $\sigma_T$, it must hold that
  $\sigma_L(\ell_0)+\sigma_R(r_1) \equiv_k \sigma_L(\ell_0)+\sigma_R(r_4)$.
  As this implies $\sigma_R(r_1)=\sigma_R(r_4)$,
  we get 
  \begin{align*}
    b = f(\ell_1,r_1 )
      =& \sigma_T( \sigma_L(\ell_1) + \sigma_R(r_1) \mod k ) \\
      =& \sigma_T( \sigma_L(\ell_1) + \sigma_R(r_4) \mod k )
      = f(\ell_1,r_4) = d,
  \end{align*}
  which is a contradiction.
  Similarly, we get that $(\sigma_L(\ell_1)+\sigma_R(r_4) \mod k) \notin F$.
  Again assuming otherwise, we have that $\sigma_R(r_3)=\sigma_R(r_4)$
  which then implies
  \begin{align*}
    c = f(\ell_0, r_3)
      =&\sigma_T( \sigma_L(\ell_0) + \sigma_R(r_3) \mod k ) \\
      =&\sigma_T( \sigma_L(\ell_0) + \sigma_R(r_4) \mod k )
      = f(\ell_0, r_4) = a,
  \end{align*}
  which is a contradiction.

  Since,
  $F \cup \{ \sigma_L(\ell_0)+\sigma_R(r_4) \mod k,
  \sigma_L(\ell_1)+\sigma_R(r_4) \mod k \} \subseteq \ZZ_k$,
  contains at least six distinct elements,
  we get $k \ge 6$ and therefore, $\cost(\Part) \geq 6$.
\end{proof}

\section{Querying a Generalized Convolution}
\label{sec:gen-ov}

In this section, we prove Theorem~\ref{thm:membership}.
The main idea is to represent the $\fquery$ problem as a matrix multiplication problem,
inspired by a graph interpretation of $\fquery$.

Let $\Dd$ be an arbitrary set and $f\from \Dd\times \Dd\to \Dd$. We assume $D$ and
$f$ are fixed throughout this section. Let $g,h\from\Dd^n\to \Mrange$ and $\bv\in
\Dd^n$ be a $\fquery$ instance.
We use $\ba\|\bb$ to denote the concatenation of $\ba\in \Dd^{m}$ and $\bb\in \Dd^{k}$.
That is
$(\ba_1,\ldots, \ba_{m})\|(\bb_1,\ldots,\bb_{k}) = (\ba_1,\ldots, \ba_{m},
\bb_1,\ldots, \bb_{k})$.
If we assume that $n$ is even,
then, for a vector $\bv \in \Dd^n$, let
$\bv^\high,\bv^{\low}\in \Dd^{n/2}$ be the unique vectors such that
$\bv^{\high} \| \bv^{\low} = \bv$.
Indeed, to achieve this assumption let $n$ be odd,
fix an arbitrary $d \in \Dd$, and
define $\widetilde{g}, \widetilde{h} \from \Dd^{n+1}\to \Mrange$ as
$\widetilde{g}(\bu_1,\ldots \bu_{n+1}) = \iverson{\bu_{n+1} = d} \cdot
g(\bu_1,\ldots \bu_{n}) $
and $\widetilde{h}(\bu_1,\ldots \bu_{n+1}) = \iverson{\bu_{n+1} = d}\cdot
h(\bu_1,\ldots \bu_{n})$ for all $\bu \in \Dd^{n+1}$.
It can be easily verified that
$(g\fconv h)(\bv) = (\widetilde{g} \fconv \widetilde{h})(\bv\| (f(d,d)))$.
Thus, we can solve the $\fquery$ instance $\widetilde{g}$, $\widetilde{h}$ and $\bv\|(f(d,d))$
and obtain the correct result.

We first provide the intuition behind the algorithm
and then formally show the existence.

\begin{figure}[t]
	\centering
	\includegraphics[width=0.5\textwidth]{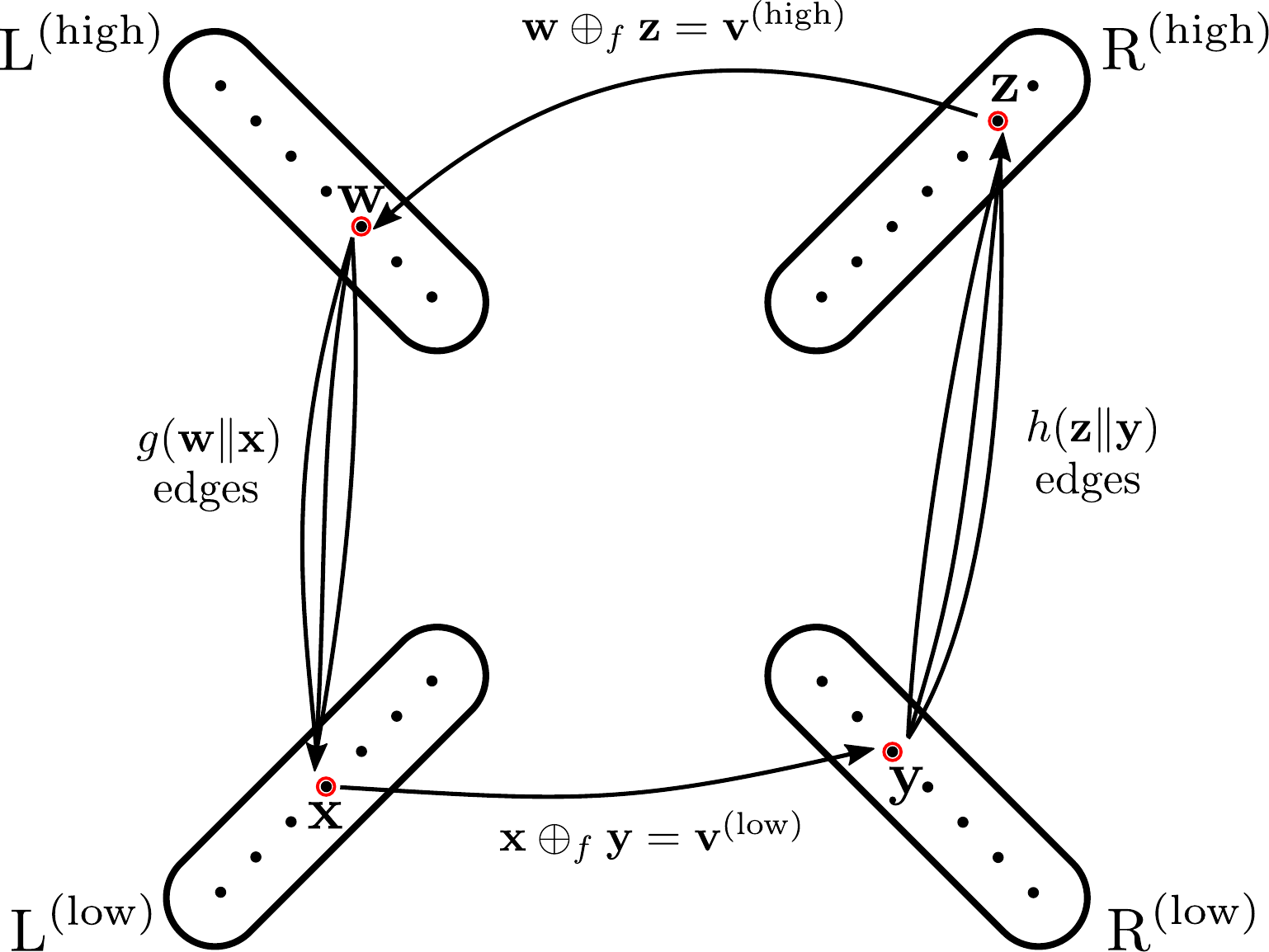}
	\caption{Construction of the directed multigraph $G$. Each
		vertex in a layer corresponds to the vector in $\Dd^{n/2}$.
		We highlighted $4$ vectors
		$\bw,\bx,\by,\bz \in \Dd^{n/2}$ each in a different layer. Note that the
		number of $4$ cycles that go through all four $\bw,\bx,\by,\bz$ is equal to
		$g(\bw\|\bx) \cdot h(\bz\| \by)$. The total
		number of directed $4$-cycles in this graph corresponds to the value
		$(g\fconv h)(\bv)$ and $\tr(W\cdot X \cdot Y\cdot Z)$.
	}
	\label{fig:ov}
\end{figure}

\subparagraph*{Intuition.}
We define a directed multigraph $G$
where the vertices are partitioned into four layers
$\Lone$, $\Ltwo$, $\Rtwo$, and $\Rone$.
Each of these sets consists of $|\Dd|^{n/2}$ vertices
representing every vector in $\Dd^{n/2}$.
For ease of notation, we use the vectors to denote the associated vertices; furthermore, the intuition assumes $g$ and $h$ are non-negative. 
The multigraph $G$ contains the following edges:
\begin{itemize}
	\item
	$g(\bw \| \bx)$ parallel edges from $\bw \in \Dd^{n/2}$ in $\Lone$
	to $\bx \in \Dd^{n/2}$ in $\Ltwo$.
	\item
	One edge from $\bx \in \Dd^{n/2}$ in $\Ltwo$
	to $\by \in \Dd^{n/2}$ in $\Rtwo$ if and only if $\bx \oplus_f \by=v^\low$.
	\item
	$h(\bz \| \by)$ parallel edges from $\by \in \Dd^{n/2}$ in $\Rtwo$
	to $\bz \in \Dd^{n/2}$ in $\Rone$.
	\item
	One edge from $\bz \in \Dd^{n/2}$ in $\Rone$
	to $\bw \in \Dd^{n/2}$ in $\Lone$ if and only if $\bw\oplus_f\bz=v^\high$.
\end{itemize}
In the formal proof,
we denote the adjacency matrix between $\Lone$ and $\Ltwo$ by $W$,
between $\Ltwo$ and $\Rtwo$ by $X$,
between $\Rtwo$ and $\Rone$ by $Y$,
and between $\Rone$ and $\Lone$ by $Z$.
See \cref{fig:ov} for an example of this construction.

\iffalse
The edges of multigraph $G$ are directed from $\Lone$ to $\Ltwo$,
from $\Ltwo$ to $\Rtwo$, from $\Rtwo$ to $\Rone$, and from $\Rone$ to $\Lone$.
The transition matrix from
$\Lone$ to $\Ltwo$ is $W$ (i.e, the number of edges between the vertex
representing $\bw\in \Dd^{n/2}$ in $\Lone$ the vertex representing  $\bx \in
\Dd^{n/2}$ in $\Ltwo$ is $W_{\bw,\bx}$). Similarly, $X$, $Y$, and $Z$ are the
transition matrices from $\Ltwo$ to $\Rtwo$, from $\Rtwo$ to $\Rone$ and from
$\Rone$ to $\Lone$ respectively. See illustration in Figure~\ref{fig:ov}.
\fi

Let $\bw, \bx, \by ,\bz \in \Dd^{n/2}$ be vertices in $\Lone$, $\Ltwo$, $\Rtwo$, and $\Rone$.
It can be observed that if $(\bw \| \bx) \oplus_f (\by\|\bz) \neq \bv$,
then $G$ does not contain any cycle of the form
$\bw \to \bx \to \by \to \bz \to \bw$
as one of the edges $(\bx, \by)$ or $(\bz, \bw)$ is not present in the graph.
Conversely, if $(\bw \| \bx)\oplus_f (\by\|\bz)= \bv$, then one can verify that
there are $g(\bw\|\bx)\cdot h(\bz\|\by)$ cycles of the form $\bw\to
\bx \to \by \to \bz \to \bw$.  We therefore expect
that $(g \fconv h)(\bv) $ is the number of cycles in $G$
that start at some $\bw \in \Dd^{n/2}$ in $\Lone$,
have length four,
and end at the same vertex $\bw$ in $\Lone$ again.

\subparagraph*{Formal Proof.}
We use the notation $\Mat_\Int(\Dd^{n/2}\times \Dd^{n/2})$
to refer to a $|\Dd|^{n/2} \times |\Dd|^{n/2}$ matrix of integers
where we use the values in $\Dd^{n/2}$ as indices.
The \emph{transition matrices} of $g$, $h$ and $\bv$ are  the matrices
$W,X,Y,Z\in \Mat_\Int(\Dd^{n/2}\times \Dd^{n/2})$  defined by
\begin{align*}
	W_{\bw,\bx} &\deff g(\bw\|\bx)
		&& \forall \bw,\bx\in \Dd^{n/2}
		\\
	X_{\bx,\by} &\deff \iverson{ \bx \oplus_f \by=\bv^\low }
		&& \forall \bx,\by\in \Dd^{n/2}
		\\
	Y_{\by,\bz} &\deff h(\bz\|\by)
		&& \forall \by,\bz\in \Dd^{n/2}
		\\
	Z_{\bz,\bw} &\deff \iverson{ \bw \oplus_f\bz=\bv^\high}
 		&& \forall \bz,\bw\in \Dd^{n/2}
\end{align*}
Recall that the \emph{trace} $\tr(A)$ of a matrix $A \in \Mat_\Int(m \times m)$
is defined as $\tr(A) \deff \sum_{i=1}^m A_{i,i}$.
The next lemma formalizes the correctness of this construction.

\begin{lemma}
	\label{lem:fquery}
	Let $n\in \Nat$ be an even number,
	$g,h \from \Dd^n\to \Int$ and $\bv\in \Dd^n$.
	Also, let $W,X,Y,Z\in \Mat_\Int(\Dd^{n/2}\times \Dd^{n/2})$
	be the transition matrices of $g$, $h$ and $\bv$.
	Then,
	\[
		(g\fconv h) (\bv) = \tr(W\cdot X \cdot Y\cdot Z).
	\]
\end{lemma}

\begin{proof}
	For any $\bw, \by\in \Dd^{n/2}$  it holds that,
	\begin{equation}
		\label{eq:WX_mult}
		(W\cdot X)_{\bw,\by} = \sum_{\bx \in \Dd^{n/2}} W_{\bw,\bx}\cdot X_{\bx, \by} =
		\sum_{ \bx\in \Dd^{n/2}} \iverson{\bx \oplus_f \by= \bv^\low} \cdot g(\bw\|\bx).
	\end{equation}
	Similarly, for any $\by,\bw\in \Dd^{n/2}$  it holds that,
	\begin{equation}
		\label{eq:YZ_mult}
		(Y\cdot Z)_{\by,\bw} = \sum_{\bz \in \Dd^{n/2}} Y_{\by,\bz}\cdot Z_{\bz, \bw} =
		\sum_{ \bz\in \Dd^{n/2}} \iverson{\bw\oplus_f\bz= \bv^\high} \cdot h(\bz\|\by).
	\end{equation}
	Therefore, for any $\bw\in \Dd^{n/2}$,
	\begin{align*}
		% \label{eq:WXYZ_mult}
		&(W\cdot X\cdot Y\cdot Z)_{\bw,\bw}
		= \sum_{\by \in \Dd^{n/2}} (W\cdot X)_{\bw,\by}\cdot (Y\cdot Z)_{\by, \bw} \\
		&= \sum_{\by \in \Dd^{n/2}}  \left(
	    \sum_{ \bx\in \Dd^{n/2}}\iverson{ \bx \oplus_f \by= \bv^\low} \cdot g(\bw\|\bx)
		\right) \left(
	    \sum_{ \bz\in \Dd^{n/2}} \iverson{ \bw\oplus_f\bz= \bv^\high} \cdot  h(\bz\|\by)
		\right) \\
	    &=\sum_{\bx,\by,\bz \in \Dd^{n/2}} \iverson{
			\bx\oplus_f\by=\bv^\low} \cdot \iverson{\bw\oplus_f\bz=\bv^\high~} \cdot g(\bw\|\bx) \cdot h(\bz\|\by) \\
		&=\sum_{\bx,\by,\bz \in \Dd^{n/2}} \iverson{
			(\bw \| \bx)\oplus_f(\bz \| \by)=\bv^\high\| \bv^\low} \cdot g(\bw\|\bx) \cdot h(\bz\|\by),
	\end{align*}
	where the second equality follows by \eqref{eq:WX_mult} and \eqref{eq:YZ_mult}.
	Thus,
	\begin{align*}
		\tr(W\cdot X~\cdot&~ Y\cdot Z ) = \sum_{\bw \in \Dd^{n/2}} (W\cdot X\cdot Y\cdot Z)_{\bw,\bw}  \\
		&=  \sum_{\bw \in \Dd^{n/2} }\sum_{~\bx,\by,\bz \in \Dd^{n/2}}
	    \iverson{ (\bw \| \bx )\oplus_f (\bz \| \by)=\bv} \cdot g(\bw\|\bx) \cdot h(\bz\|\by) \\
		& = \sum_{\bu, \bt \in \Dd^{n}}  \iverson{\bu\oplus_f \bt = \bv} \cdot  g(\bu)\cdot h(\bt) \\
		&= (g\fconv h) (\bv).
		\qedhere
	\end{align*}
\end{proof}

Now we have everything ready to give the algorithm for \fquery.
\begin{proof}[Proof of Theorem~\ref{thm:membership}]
	The algorithm for solving \fquery
	works in two steps:
	\begin{enumerate}
		\item
		\label{fquery:transition}
		Compute the transition matrices $W$, $X$, $Y$, and $Z$
		of $g$, $h$ and $\bv$ as described above.
		\item
		\label{fquery:mult}
		Compute and return $\tr(W \cdot X \cdot Y \cdot Z)$.
	\end{enumerate}
	By \cref{lem:fquery} this algorithm returns
	$(g\fconv h)(\bv)$. Computing the transition matrices in
	Step~\ref{fquery:transition} requires $\Oc{|\Dd|^n}$ time. Observe the maximal absolute values of an entry in the transition matrices is $M$. The computation of
	$W\cdot X\cdot Y\cdot Z$ in Step~\ref{fquery:mult} requires three matrix
	multiplications of $|\Dd|^{n/2}\times |\Dd|^{n/2}$ matrices, which can be done
	in $\Oc{(|\Dd|^{n /2})^\omega}$ time.
	Thus, the overall running time of the
	algorithm is $\Oc{|\Dd|^{\omega \cdot n / 2}}$.
\end{proof}

\section{Conclusion and Future Work}
\label{sec:conclusion}

In this paper, we studied the \fConv problem and demonstrated
that the naive brute-force algorithm can be improved for every $f \from D \times
D \to D$. We achieve that by
introducing a \emph{cyclic partition} of a function and showing that there always
exists a cyclic partition of bounded cost.
We give an $\Oc{(c|\Dd|^2)^{n}}$ time
algorithm that computes \fConv for $c \deff
3/4$ when $|\Dd|$ is even. 

The cyclic partition is a very general tool and potentially it can be used to
achieve greater improvements for certain functions $f$. For example, in
multiple applications (e.g.,~\cite{cut-and-count,csr21,cut-count2,cut-count3}) the function
$f$ has a cyclic partition with a single cyclic minor. Nevertheless, in our proof we
only use cyclic minors where one domain is of size is at most $2$. 
We suspect that larger minors have to be considered to obtain better results.
Indeed, the lower bound from \cref{lem:existenceOfMinor:tightness} implies
that our technique of considering two \emph{arbitrary} rows together
cannot give a faster algorithm than $\Oc{(3/4 \cdot \abs{\Dd}^2)^n}$ in general.
An improved algorithm would have to select these rows very carefully
or consider three or more rows at the same time.

We leave several open problems. Our algorithm offers an exponential (in $n$) improvement
over a naive algorithm for domains $\Dd$ of constant size. Can we hope for an
$\Oc{|\Dd|^{(2-\eps)n}}$ time algorithm for \fConv for some
$\eps > 0$? We are not aware of any lower bounds, so in principle even
an
$\Oc{|\Dd|^n}$ time algorithm is plausible. 

\begin{figure}[t]
    \centering
	\includegraphics[width=0.9\textwidth]{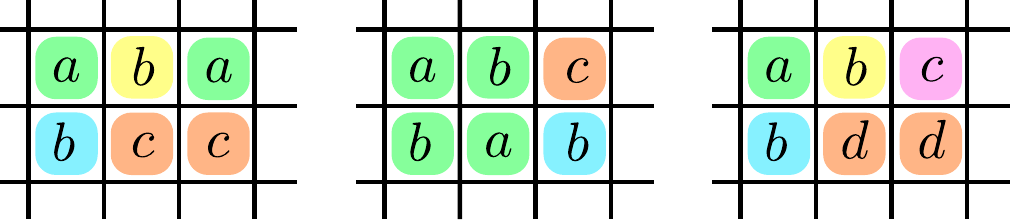}
    \caption{Here are three concrete examples of functions $f$ for which we
    expect that the running times for \fConv should be $\Oc{3^n}$,
$\Oc{3^n}$ and $\Oc{4^n}$. However, the best cyclic partitions for this
functions have costs $4$, $4$ and $5$ (the partitions are highlighted
appropriately). This implies that the best running time, which may be
attained using our techniques are $\Oc{4^n}$, $\Oc{4^n}$ and $\Oc{5^n}$. }
    \label{fig:open}
  \end{figure}

Ideally, we would expect that the \fConv problem can be solved in
$\Oc{(|L|^n+|R|^n+|T|^n)}$ for any function $f \from L \times R \to T$.
In Figure~\ref{fig:open} we
include three examples of functions that are especially difficult for our
methods.

Finally, we gave an $\Oc{|\Dd|^{\omega \cdot n / 2}}$ time algorithm for
\fquery problem. For $\omega = 2$ this algorithm runs in almost
linear-time, however for the current bound $\omega < 2.372$ our algorithm
runs in time $\Oc{|\Dd|^{1.19n}}$. Can $\fquery$ be solved in
$\Oc{|\Dd|^n}$ time without assuming $\omega=2$?

\begin{picture}(0,0)
\put(350,-30)
{\hbox{\includegraphics[width=40px]{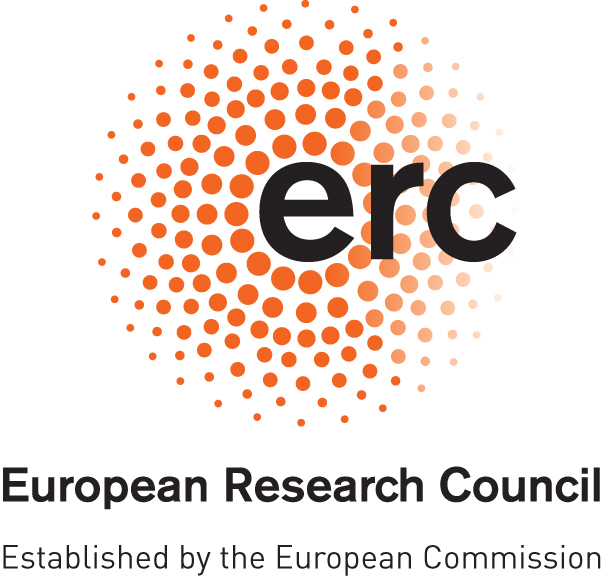}}}
\put(295,-50)
{\hbox{\includegraphics[width=60px]{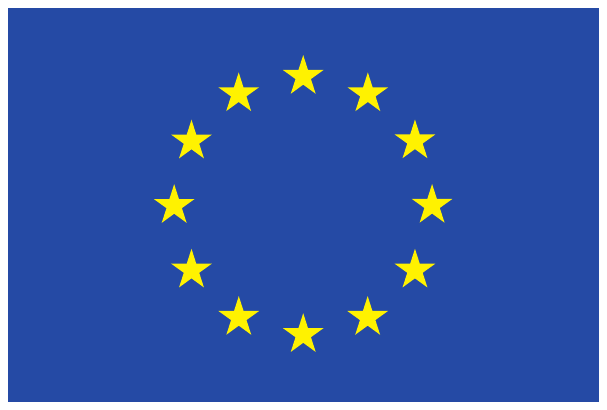}}}
\end{picture}

\bibliographystyle{plain}
\bibliography{bib}

\begin{appendix}
    \section{Proof of Theorem~\ref{thm:cyclic}}%
\label{sec:cyclic-convolution}

In this section we prove Theorem~\ref{thm:cyclic}. We crucially rely on the
following result by van Rooij~\cite{Rooij20}.

\begin{theorem}[{\cite[Lemma 3]{Rooij20}}]
	\label{thm:alg-convolution}
	There is an algorithm which given $k\in \mathbb{N}$, $\br\in \mathbb{N}^k$ a prime $p$, an $\br_j$-th primitive root of unity $\omega_j$ for every $j\in [k]$ and two functions 	$g,h\from \Int_{\br_1} \times \dots \times \Int_{\br_k} \to \Int$ computes the cyclic convolution of $g$ and $h$ modulo $p$ (that is, return a function $\phi$ such that
	$\phi(\bq)=(g \odot h)(\bv) \mod p$ for every
	$\bv \in \Int_{r_1}\times \dots \times \Int_{r_k}$)
	in $\Oh( R\log (R))$ arithmetic operations where $R=\prod_{j=1}^k \br_j$. 

\end{theorem}

Ideally, we would like to use the algorithm from \Cref{thm:alg-convolution} with
a sufficiently large prime $p$ such that the values of $g \conv h$ could be recovered from the values of $g \conv h$  modulo $p$. Finding such a prime $p$ along with the required roots of unity is, however, a non trivial task which we do not know how to perform deterministically while retaining the running time at $\Occ{R}$.
The basic idea behind our approach is to compute $g \odot h$ modulo
$p_i$ for a sufficiently large number of distinct small primes $p_i$ using \Cref{thm:alg-convolution}.  If
$\prod_i p_i $ is sufficiently large, then the values of $g \odot h$  can be uniquely recovered using the
Chinese Remainder Theorem.

\begin{theorem}[Chinese Remainder Theorem]
	\label{thm:chinese}
	Let $p_1,\dots,p_m$ denote a sequence of integers that are pairwise coprime and define $P
	\deff \prod_{i \in [m]}p_i$.
	Also let $0 \leq a_i < p_i$ for all $i \in [m]$.
	Then there is a unique number $0 \leq s < P$ such that
	\[s \equiv a_i \mod {p_i}\]
	for all $i \in [m]$.
	Moreover, there is an algorithm that, given $p_1,\dots,p_m$ and
	$a_1,\dots,a_m$, computes the number $s$ in time $\Oh((\log P)^2)$.
\end{theorem}

To find the small primes for the application of the Chinese Remainder Theorem,
we additionally use density properties of primes in arithmetic progression. Given $q\in
\mathbb{N}$, we say $p\in \mathbb{N}$ is a {\em $q$-prime} if $p$ is a prime
number and $p\equiv 1 \mod q$.
We use $\prime_q(i)$ to denote the $i$-th $q$-prime.
That is, $\prime_q(i)$ is a $q$-prime
such that the number of $q$-primes smaller than $\prime_q(i)$ is exactly $i-1$.
Also, for any $B,q\in \mathbb{N}$, we define
\begin{equation*}
	\primebound_q(B) \deff \min\left\{ m\in \mathbb{N}~\middle|~\prod_{i=1}^{m} \prime_q(i) \geq B \right\}\end{equation*}
to be the minimal number $m$ such that the product of the first $m$ $q$-primes is at least $B$.  We use the following upper bound on $\primebound$. 

\begin{lemma}
	\label{lem:prime_bound}
	Let $B, q \in \mathbb{N}$ be integers such that $B,q\geq 3$ 
  and $m=\textnormal{\primebound}_q(B)$. Then $m\leq \ln (B)+1$ and $\textnormal{\prime}_q(m)
  \leq\max\left\{ \exp\left(8\cdot \sqrt{q} \cdot \ln^3(q) \right),~ \exp(q) , ~2q\cdot \ln(B)  \right\}$. 
\end{lemma} 
In the proof of \Cref{lem:prime_bound} we use a known result for the density of primes in arithmetic progressions taken from \cite{BennettMOR18}. For any $x,q \in \mathbb{N}$, define $\theta(x,q)$ to be the sum of $\ln(p)$ for all $q$-primes $p$ such that $p\leq x$. Formally, we define
\begin{equation*}
%	\label{eq:theta_def}
	\theta(x,q) \deff \sum_{i=1}^{\infty} \iverson{\prime_q(i) \leq x } \cdot \ln \left(\prime_q(i)\right).
\end{equation*}
With this definition,
we can now state the result about the density
of primes in arithmetic progressions.
\begin{lemma}[{\cite[Corollary 1.8]{BennettMOR18}}]
	\label{lem:density_prob}
	Let $q$ and $x$ be integers with
	$q>3$ and $x\geq \exp(8\cdot \sqrt{q}\cdot \ln^3q)$.
	Then,
	\[
		\theta(x,q) \geq \frac{x}{\varphi(q)} - \frac{1}{160}\cdot \frac{x}{\ln x}
	\]
	where $\varphi$ is Euler's totient function.
\end{lemma}
Now we have everything ready to prove \cref{lem:prime_bound}.
\begin{proof}[Proof of \Cref{lem:prime_bound}]
	We first prove the bound for $m$.
	By the definition of $m$ as $m=\primebound_q(B)$,
  we get
	$\prod_{i=1}^{m-1} \prime_q(i) <B$.
  As $\ln(\prime_q(i))>1$ for every $i$,
  we have
  \begin{equation*}
    m-1 < \sum_{i=1}^{m-1}  \ln(\prime_q(i))  = \ln \left(\prod_{i=1}^{m-1} \prime_q(i) \right) <\ln(B)
  \end{equation*}
  which implies $m<\ln(B)+1$.

Now we prove the bound for $\prime_q(m)$.
For this we set
\[
	x= \max\left\{ \exp\left(8\cdot \sqrt{q} \cdot \ln^3(q) \right),~ \exp(q) , ~2q\cdot \ln(B)  \right\}.
\]
By \cref{lem:density_prob}, we get
\begin{align*}
	\theta(x,q)   &\geq  \frac{x}{\varphi(q)} - \frac{1}{160} \cdot \frac{x}{\ln x }\\
	&=  x \left( \frac{1}{\varphi(q)} - \frac{1}{160 \cdot \ln x}\right)
\end{align*}
and, using \( \varphi(q)\leq q \) and \( \ln(x)\geq  q \), we have
\begin{align}
				\label{eq:theta_prop}
	\theta(x,q)
	&\geq x \cdot \left( \frac{1}{q } - \frac{1}{160 \cdot q} \right)  \nonumber\\
	&\geq x \cdot \frac{1}{2\cdot q } \nonumber\\
	&\geq  \ln B.
\end{align}

% The third inequality holds as $\varphi(q)\leq q$ and $\ln(x)\geq  q$. 

Let $\ell=\max \{j ~|~\prime_q(j)\leq x \}$ be the index of the largest $q$-prime which is  not greater than  $x$. Then,
\begin{align*}
\prod_{i=1}^{\ell} \prime_q(i)   &= \exp \left( \sum_{i=1}^{\ell} \ln(\prime_q(i))\right) \\
&= \exp \left( \sum_{i=1}^{\infty} \iverson{\prime_q(i)\leq x }\cdot \ln(\prime_q(i))\right) \\
&= \exp \left(\theta(x,q)\right) \geq B,
\end{align*}
where the  inequality follows from \eqref{eq:theta_prop}.
By the definition of $\primebound_q$,
we get $m=\primebound_q(B)\leq \ell$.
Hence, $\prime_q(m )\leq \prime_{q}(\ell) \leq x $
which finishes the proof.
	\end{proof} 

In the remainder we give the $\Occ{(\prod_{i=1}^k \br_i)}$ algorithm for the
$K$-\CyConvProb.

\begin{proof}[Proof of \Cref{thm:cyclic}]
Fix a finite set $K=\{c_1,\dots, c_{\ell}\}\subseteq \Nat$
which is considered as a constant throughout this proof.
Let integers $k, M \in \Nat$, integer vector $\br \in K^k$
and functions $g,h\from Z \rightarrow \Mrange$ where
$Z=\Int_{\br_1}\times \dots \times \Int_{\br_k}$
be an input for the $K$-\CyConvProb.

For every $t\in [\ell]$, let $D_t$ be the prime factors of $c_t$.
We define $R= \prod_{j=1}^{k} \br_j$ and  
observe that for any~$\bv\in Z$ it holds that $\abs{ (g \conv h) (\bv)} \leq
R\cdot    M^2 $. 
Further define $B \deff   3 \cdot  R \cdot M^2 $ and $q= \prod_{c\in K}
c = \prod_{t=1}^{\ell} c_t$.
Assume without loss of generality that $q\geq 3$ and  note that $q$
depends only on the fixed finite set $K$
and therefore, can be viewed as a constant.

With this notation we can formally state the algorithm.
\begin{enumerate}
	\item
	Iterate over the numbers of the form $q\cdot a+1$
	for $a\in \{1,2,\ldots\}$
	and test for each one if it is prime.
	The process continues until the
	product of the $q$-primes exceeds $B$.
	Denote these numbers by $p_1,\dots,p_m$.

	\item
	\label{primolution:iterate}
	For every $i\in [m]$ and $t\in [\ell]$,
	iterate over all elements $x \in \FF_{p_i}$
	and test whether $x^{c_t}\equiv 1 \mod p_i$
	and $x^{{c_t}/{d}} \not\equiv 1 \mod p_i$
  for every $d\in D_t$.
	If so, then set $x$ as the $c_t$-th root of unity in $\FF_{p_i}$.

	\item
	For all $i\in [m]$,
	use \cref{thm:alg-convolution} with the prime $p_i$
	and appropriate roots of unity
	to compute the function $f^{(i)}\from Z\to \mathbb{Z}_{p_i}$
	defined by
	\[
		f^{(i)}(\bv)\deff (g \odot h)(\bv) \mod{p_i}~~~~~~\forall \bv\in Z
		.
	\]

	\item
	Define $P=\prod_{i=1}^{m} p_i$, we define a function  $f_P\from Z\to \mathbb{Z}_{P}$ as follows. 
	For each $\bv \in Z$,
	use the Chinese Remainder Theorem (cf.~\cref{thm:chinese}) to compute
	the value $0\leq f_P(\bv)<P$
	such that $f_P(\bv) \equiv f^{(i)}(\bv) \mod p_i$ for all $i\in [m]$.
	% from the values $f^{(1)}(\bv),\dots,f^{(m)}(\bv)$
	%with the primes $p_1,\dots,p_m$.
	%Then, define the function $f_P\from Z\to \mathbb{Z}_{P}$
%	with $\bv \mapsto a(\bv)$.
	\item Finally,
	compute the function $f\from Z\to \mathbb{Z}$ using the formula
	\[
	f(\bv)=
	\begin{cases} 
		f_P(\bv)  &\textnormal{ if }f_P(\bv) <\frac{P}{2} \\
		f_P(\bv) - P & \textnormal{ if } f_P(\bv) \geq \frac{P}{2}
	\end{cases} \] 
	for all $\bv\in Z$ and return $f$.
\end{enumerate}
Before we move to proving the correctness,
we first argue that the algorithm is well-defined.
From the definition,
the first step computes the first $m=\primebound_q(B)$ $q$-primes
such that $p_1 = \prime_q(1), \ldots, p_m=\prime_q(m)$.
It remains to show that,
for every $i \in [m]$ and $t \in [\ell]$,
the $c_t$-th primitive root of unity in $\FF_{p_i}$ exists.
Indeed, since $c_t$ divides $p_i - 1$
(which is in turn true as $p_i \equiv 1 \mod q$ and $c_t$ divides $q$),
such a root of unity exists.
Moreover, as $D_t$ contains all prime factors of $c_t$,
one can easily show that it actually suffices to consider
only values of the form $x^{c_t/d}$ for every $d\in D_t$
to correctly decide if $x$ is a primitive $c_t$-th root of unity in $\FF_{p_i}$.
The application of \cref{thm:alg-convolution} in the second step is possible
as $\br_j \in K=\{c_1,\ldots, c_{\ell}\}$ for every $j\in [n]$
and the roots of unity are computed by the second step.

Now we argue about the correctness of the algorithm.
\begin{claim}
	\label{clm:cyclic:correctness}
	For all $\bv \in Z$, we have $f(\bv) = (g \conv h)(\bv)$.
\end{claim}
\begin{claimproof}
	As the algorithm is well defined,
	the third step computes,
	the convolution of $g$ and $h$ modulo $p_i$ for every $i \in [m]$.
	
	Now fix some $\bv \in Z$.
	We define $b(\bv) = (h\conv g)(\bv) \mod P$
	and observe $0\leq b(\bv) < P$.
	Moreover, for every $i\in [m]$ it holds that
	\[ b(\bv) \mod p_i
		= \left((h\conv g)(\bv) \mod P\right) \mod p_i
		= (h\conv g)(\bv) \mod p_i
		= f^{(i)}(\bv).
	\]
	Since \cref{thm:chinese} also guarantees the resulting number to be unique,
	it follows that $f_P(\bv) = b(\bv)$
	which implies $f_P(\bv) = (g \conv h )(\bv) \mod P$.

	Now we focus on the last step.
	By the definition of $m = \primebound_q(B)$,
	it holds that $P= \prod_{i=1}^{m}p_i \geq B = 3 \cdot R\cdot M^2$.
	Consider the following cases.
	\begin{itemize}
	\item 
	 In case $(g\conv h) (\bv) \geq 0$ we have 
	\[(g\conv h) (\bv)\leq R\cdot M^2  < \frac{B}{2}\leq P.\]
	This implies that $f_P(\bv) = (g\conv h)(\bv) \mod P
	= (g\conv h)(\bv) <\frac{B}{2}$.
	Thus, $f(\bv) = f_P(\bv) = (g\conv h)(\bv)$.

	\item In case $(g\conv h) (\bv) < 0$
	it holds that  \[(g\conv h) (\bv)\geq -R\cdot M^2  >-P.\]
	This now implies that
	\[f_P(\bv) = (g\conv h)(\bv) + P\geq  P- R\cdot M^2 > \frac{P}{2}.\]
	Hence, $f(\bv) = f_P(\bv) - P = (g\conv h)(\bv) + P- P = (g\conv h)(\bv)$.
	\end{itemize}
Hence, $f(\bv) =(g \conv h) (\bv )$ for all $\bv \in Z$,  which concludes the proof.
\end{claimproof}
From \cref{clm:cyclic:correctness} we know that the algorithm is correct
and the function $f$ returned by the algorithm is indeed $(g\conv h)$.
It only remains to analyze the running time of the procedure.
\begin{claim}
	\label{clm:cycle:runtime}
	The procedure terminates in time $\Occ{R}$.
\end{claim}
\begin{claimproof}
	We consider each step on its own.
	\begin{enumerate}
		\item
		Since prime testing can be
		done in polynomial time (in the representation size of the number), we can find the sequence $p_1,\dots,p_m$ in time $\Oh(p_m\cdot \polylog  p_m)$. By \Cref{lem:density_prob}, and since $q$ is a constant,  it follows that
		\[
			p_m \leq
				\max\left\{
					\exp\left(8\cdot \sqrt{q} \cdot \ln^3(q) \right),\,
					\exp(q),\,
					2q\cdot \ln(B)
				\right\}
			= \Oh(\ln(B))
			= \Oh(\log (R\cdot M))
		\]
		and $m\leq \ln(B)+1 = \ln(3RM^2)+1$.  Hence, the running time of this  step is 
		$\Oh(p_m\cdot \polylog  p_m)= \Oh(\polylog(R\cdot M))$.

		\item
		For each $i\in[m]$ and $t\in [\ell]$, in Step~\ref{primolution:iterate} of the algorithm 
		we iterate over $p_i$ values and check $\abs{D_t}$ values.
		Since $D_t$ are the prime factors of $c_t$ (and hence $|D_t|$ is a constant),
		this takes time $\Oh(p_i \polylog p_i)$
		which can be bounded by $\Oh(\polylog(R\cdot M))$.

		Since $m\leq \log(3\cdot R\cdot M^2)+1$ and $\ell$ is a constant,
		the overall running time of the step is $\Oh(\polylog(R\cdot M))$.

		\item
		By \cref{thm:alg-convolution}, the number
		of arithmetic operations required to compute $f^{(i)}$ is
		$\Oh(R\cdot\log(R))$. Since each arithmetic operation is performed in
		$\FF_{p_i}$, the total time spent to compute $f^{(i)}$ is
		\[
			\Oh(R\cdot\log(R)\cdot \log^2(p_i))
			= \Oh(R\cdot\log(R)\cdot \log^2( \log (R \cdot M ) ))
			= \Oh(R\cdot \polylog(R\cdot M)),
		\]
		where the first equality holds
		because $p_i\leq p_m=\Oh(\log(R\cdot M))$.

		Finally, as $m=\Oh(\log(R\cdot M))$,
		the overall computation time of this step is
		$m\cdot \Oh(R\cdot \polylog(R\cdot M) =\Oh(R\cdot \polylog(R\cdot M))$.
		
		\item
		As we iterate over all $R$ values from $Z$ and by \cref{thm:chinese},
		this computation can be done in time $\Oh(R\cdot (\log P)^2)$.
		Since $\log P \leq m \cdot p_m \leq
		% %\log (R\cdot M )\cdot \Oh(\log(R\cdot M)) =
		\Oh(\polylog(R\cdot M))$
		the overall running time of this step is $\Oh(R\cdot \polylog (R\cdot M))$. 

		\item
		As we again iterate over all elements from $Z$,
		the computation time of this step is
		$\Oh(R\cdot \polylog P)= \Oh(R\cdot \polylog (R\cdot M))$
		where we use $P=\Oh(\polylog(R\cdot M))$.
	\end{enumerate}

	As the running time of each step is at most $\Occ{R}$,
	the overall running time of the algorithm is $\Occ{R}$.
\end{claimproof}
The proof now follows by \cref{clm:cyclic:correctness,clm:cycle:runtime}.
\end{proof}

\end{appendix}

\end{document}